\documentstyle[preprint,prd,aps,psfig,floats,epsf]{revtex}
\newcommand{\be}{\begin{equation}}
\newcommand{\ee}{\end{equation}}
\newcommand{\bea}{\begin{eqnarray}}
\newcommand{\eea}{\end{eqnarray}}

\renewcommand{\L}{{\Lambda_{QCD}}}
\newcounter{dafigcounter}
\newcommand{\pfig}[3]{
 \refstepcounter{dafigcounter}
 \begin{minipage}[t]{#2}
  \begin{center}
   {\epsfxsize=#2 \mbox{\epsffile{#1.eps}}}
  \end{center}
  \label{#1}
  \small \bf Fig.~\thedafigcounter\rm\ #3
 \end{minipage}
}
\begin{document}
\draft
\tighten
\firstfigfalse
%\twocolumn[\hsize\textwidth\columnwidth\hsize\csname@twocolumnfalse\endcsname
%
\title{Phase Space Description of the Leading Order 
Quark and Gluon Production from a Space-Time Dependent Chromofield}
\author{Dennis D. Dietrich, Gouranga. C. Nayak, and Walter Greiner}
\address
{\small\it{Institut f\"ur Theoretische Physik,
J. W. Goethe-Universit\"at,
60054 Frankfurt am Main, Germany}}
%
%\date{\today} 
%{\tiny\verb$Id: sig.tex,v 2.4 1999/03/08 23:29:03 misha Exp $}}
\maketitle    

%%%%%%%%%%%%%%%%%%%%%%%%%%%%%%%%%%%%%%%%%%%%%%%%%%%%%%%%%%%%%%%%%%%%%%%%%%%%%%%
\begin{abstract}

We derive source terms for the production of quarks and gluons from the QCD 
vacuum in the presence of a space-time dependent external chromofield 
$A_{cl}$ to the order of $S^{(1)}$. We found that the source 
terms for the parton production processes $A_{cl} \rightarrow q\bar{q}$ 
and $A_{cl},A_{cl}A_{cl} \rightarrow gg$ also include the annihilation
processes $q\bar{q} \rightarrow A_{cl}$ and 
$gg \rightarrow A_{cl},A_{cl}A_{cl}$. 
The source terms we derive are applicable for the description of the 
production of partons with momentum $p$ larger rhan $gA$ which itself must be 
larger than $\L$. We observe that these source terms for the production of 
partons from a space-time dependent chromofield can be used to study the 
production and equilibration of the quark-gluon plasma during the very early 
stages of an ultrarelativistic heavy-ion collision.

\end{abstract}

\bigskip

%
%\narrowtext

%%%%%%%%%%%%%%%%%%%%%%%%%%%%%%%%%%%%%%%%%%%%%%%%%%%%%%%%%%%%%%%%%%%%%%%%%%%%%%%

\section{Introduction}

Recently, a lot of effort is made to study the production and equilibration
of the quark-gluon plasma in ultrarelativistic 
heavy-ion collisions at RHIC and LHC \cite{qgp}.
Such a state of matter is predicted by lattice QCD calculations at high 
temperatures and high densities \cite{lattice}.
The major problems one encounters in such a study is how quarks and gluons are 
formed in these experiments and how their phase-space 
distribution evolves in space-time 
with collisions among the partons taken into account. At ultrarelativistic 
energies, the two nuclei are highly Lorentz contracted. When they pass through 
each other, a chromoelectric field is formed due to the exchange of soft 
gluons 
\cite{soft,alll,nayak,sr,more}. This is a natural extension of the color 
flux-tube or
the string model which are widely applied to high energy pp, $e^+e^-$ and pA 
collisions \cite{coll,lund}. This chromoelectric field
polarizes the QCD vacuum which leads to the production of
quark/antiquark pairs and gluons via a Schwinger-like mechanism. 
These quarks and gluons collide with each other to
form a thermalized quark-gluon plasma. The space-time evolution
of the formed partons can be studied by solving the relativistic
non-abelian transport equations for quarks and gluons \cite{nayak,heinz}. 
As the chromofield exchanges color with quarks and gluons, color
is a dynamic quantity. The time evolution of the classical color charge 
follows Wong's equations \cite{wong}

\be
\frac{dQ^a}{d\tau}=f^{abc}u_{\mu}Q^bA^{c\mu}
\ee

and

\be
\frac{dp^\mu}{d\tau}=Q^aF^{a\mu\nu}u_\nu.
\ee

Here, the first equation describes the precession of the color charge
in the presence of a classical background chromofield. The second equation
is the non-abelian version of the Lorentz-force equation. Taking
Wong's equations into account, one finds the relativistic non-abelian
transport equation \cite{nayak,heinz} which for quarks reads for example as:

\be
\left[ p_{\mu} \partial^\mu + g Q^a F_{\mu\nu}^a p^\nu 
\partial^\mu_p
+ g f^{abc} Q^a A^b_\mu
p^{\mu} \partial_Q^c \right]  f(x,p,Q)=C+S.
\label{trans1}
\ee

Note that there are different transport equations for quarks, 
antiquarks, and gluons. In the equations mentioned above, 
$f(x,p,Q)$ is the single-particle distribution function of quarks, antiquarks, 
and gluons respectively in the 14-dimensional extended phase space of 
coordinate, momentum 
and color in the $SU(3)$ gauge group. The first term on the LHS corresponds 
to common convective flow, the second term is
the non-abelian generalization of the Lorentz force term and the third term
represents the precession of the 
color charge in the presence of the background field
as described by Wong's equations. On the 
RHS, $C$ denotes the collision term which describes the collisions among 
the partons and $S$ represents the source term for the production of the 
respective particle species. The source term $S$ contains the detailed 
information about the creation of 
partons from the chromofield. It is defined as the probability 
for the production of a parton per unit time 
per unit volume of the phase space. For this 
reason, the development of the QGP is fundamentally governed by this 
source term $S$.

The formation and equilibration of the quark-gluon plasma within the color
flux-tube model is studied by several authors \cite{alll,nayak}. However,
in all these studies, source terms for parton production from a constant
chromoelectric field were employed, because source terms for parton production
from a space-time dependent chromofield were not studied in literature
before. In fact, the chromofield acquires a strong 
space-time dependence as soon as it starts producing partons. Also
the acceleration, collision and precession terms present in the
transport equation make the field space-time dependent. This implies
that the source term for particle production from a constant field 
is not applicable under these circumstances. 

We mention here that particle production from the classical field
via vacuum polarization is studied in two different cases. 
In the presence of a constant uniform background field
the particle production is computed through the Schwinger mechanism 
\cite{schwinger} which is an exact one-loop non-perturbative result. 
This result can also be understood as semi-classical tunneling 
across the mass gap \cite{casher}. However, for a space-time
dependent field, particles can be produced directly by a perturbative 
mechanism \cite{schwinger,izju}.
It is observed in numerical studies \cite{nayak} that the chromofield acquires 
a strong space-time dependence as soon as it produces partons.
Due to these reasons, formulas based on 
constant fields are not applicable to such problems and source terms
for parton production from a space-time dependent chromofield are
necessary. In this paper, we will derive the source terms for the creation of
$q\bar{q}$ and $gg$ pairs from a space-time dependent chromofield in order of 
$S^{(1)}$.

The production of $q\bar{q}$ pairs from a space-time dependent 
non-abelian field is similar to the 
creation of $e^+e^-$ pairs from an abelian field. 
Apart from the color factors, the interaction 
of a quantized Dirac field with the classical potential is the same in both 
cases. So, the results obtained for the probability of creation of $e^+e^-$ 
pairs from a Maxwell field can be transfered to the creation of 
$q\bar{q}$ pairs from a Yang-Mills field.
Contrary to that, the production of gluons from a classical external 
Yang-Mills field has no analogue in the abelian case. There exist direct
interactions between the quantized non-abelian field and the classical 
non-abelian potential which then lead to the production of gluons from the 
QCD vacuum.
In this paper, we will derive the source term for the production of gluons 
from a space-time dependent chromofield through the application of the 
background field method of QCD which was developed by DeWitt \cite{dewitt} and 
later on extended by 't Hooft \cite{thooft}.

As already mentioned before, in the color flux-tube model the creation of the 
chromofield 
is via soft gluon exchange when two nuclei cross each other 
\cite{soft,alll,nayak,sr,more}. It is the extension of the string model widely 
applied to high-energy $e^+e^-$-, $pp$- and $pA$- collisions \cite{coll,lund}. 
However, this is not the only model where a chromofield is created in high 
energy heavy-ion collisions. The McLerran-Venugopalan model also predicts the
existence of a chromofield at low x and low transverse momentum 
\cite{mclerran,raju}. 
Also recent work by Makhlin and Surdotovich \cite{maksur} and many others,
 {\it e.g.} \cite{jana,bodeker} seems to indicate that the highly 
gluon dominated environment in heavy-ion collisions will be ruled by 
scales resulting from the presence of background fields. In any
case, once there is a chromofield it will polarize the QCD vacuum and will
produce quarks and gluons which will then form a thermalized quark-gluon
plasma. In the McLerran-Venugopalan model the parton production is 
described as classical radiation whereas we describe parton production via
vacuum polarization.
It might be interesting to incorporate gluon creation both by classical 
radiation 
and vacuum-polarization in the transport equation to study the evolution of 
the quark-gluon plasma.

The paper is organized as follows: In chapters II and III we present the 
source terms 
for quark and gluon production from a space-time dependent chromofield. In 
chapter IV, we analyze the production of partons from a purely time dependent 
background field. Finally, we summarize and discuss our results in chapter V 
and  conclude in chapter VI. There, one can also find a treatment on the 
applicability of the source terms presented in this paper. In the end, there 
follows an Appendix incorporating more mathematical details than were 
suitable for the main body of the paper.

%%%%%%%%%%%%%%%%%%%%%%%%%%%%%%%%%%%%%%%%%%%%%%%%%%%%%%%%%%%%%%%%%%%%%%%%%%%%%%%

\section{Source Term for Quark Production from a Space-Time Dependent 
Chromofield}

The amplitude for the lowest order process $A\rightarrow q(k_1)\bar{q}(k_2)$ 
(see Fig. (1))
contributing to the production of $q\bar{q}$ pairs from a space-time 
dependent classical non-abelian field $A^{a\mu}$ via vacuum-polarization is 
given by:

\bea
M=<q(k_1)\bar{q}(k_2)|S^{(1)}|0>
%~\nonumber \\
=
\bar{u}^i(k_1)(V^F)_{ij\mu}^aA^{a\mu}(K)v^j(k_2)
\label{qamp}
\eea

with the interaction vertex:

\be
(V^F)_{ij\mu}^a=ig\gamma_{\mu}T^a_{ij}.
\ee

Here $k_1$ and $k_2$ stand for the four momenta of the outgoing quark and 
antiquark respectively and $A^{a\mu}(K)$ is the Fourier transform 
of the space-time dependent chromofield $A^{a\mu}(x)$ with $K=k_1+k_2$:

\be
A^a_{\mu}(K)=
\frac{1}{(2\pi)^2}\int d^4x A^a_{\mu}(x) ~ e^{+iK\cdot x}.
\label{fourier}
\ee
 
$\bar{u}^i$ and $v^j$ represent 
the Dirac spinors of the outgoing quark and the outgoing antiquark, 
respectively. Note that in the end the integrations over $K$ have to be 
carried out over 
the correct kinematical region, because real quarks are to be produced. It is 
given by $(K)^2>4m^2$ and $K^0>0$. This fact is not explicitly mentioned in 
the formulas, but has got to be kept in mind. 

In order to obtain the probability, the absolute square of the amplitude has 
got to be integrated over the phase space of the outgoing particles:

\be
W= 
\sum_{spin}
\int
\frac{d^3k_1}{{(2\pi)}^3 2 k_1^0}
\frac{d^3k_2}{{(2\pi)}^3 2 k_2^0}
MM^*.
\label{prob}
\ee

Putting in the definitions for the amplitudes and performing the spin-sum, as
well as carrying out one of the $d^4k$ integrations yields \cite{izju}:

\bea
W_{q\bar{q}}^{(1)}=
\frac{g^2}{4 \pi^2} 
\int d^4K 
\int \frac{d^3\vec k_1}{2 \omega_1} \frac{d^3\vec k_2}{2 \omega_2} 
\delta^{(4)}(K-k_1-k_2) 
~A^a_{\mu}(K)A^a_{\nu}(-K) 
~\nonumber \\
{[k^{\mu}_1 k^{\nu}_2+k^{\mu}_2 k^{\nu}_1 -\frac{K^2}{2}g^{\mu\nu}]}
\label{wqq}
\eea

where $\omega_{1,2}$ is defined as:

\be
\omega_{1,2}=\sqrt{|\vec k_{1,2}|^2+m^2}.
\label{om}
\ee

Using the Fourier transform Eq.(\ref{fourier}) of the classical field in 
Eq.(\ref{wqq}) one finds:

\bea
W_{q\bar{q}}^{(1)}=
\frac{g^2}{(2\pi)^6} 
\int d^4K 
\int \frac{d^3\vec k_1}{2 \omega_1} \frac{d^3\vec k_2}{2 \omega_2} 
\delta^{(4)}(K-k_1-k_2)
\int d^4x_1d^4x_2
~\nonumber \\ 
e^{iK(x_1-x_2)}A^a_{\mu}(x_1)A^a_{\nu}(x_2) 
{[k^{\mu}_1 k^{\nu}_2+k^{\mu}_2 k^{\nu}_1 -\frac{K^2}{2}g^{\mu\nu}]}.
\label{wqqf}
\eea

From the resulting expression (\ref{wqqf}), it is then possible to extract the 
gauge invariant source term for the production of $q\bar{q}$ pairs. It should
be mentioned here that the probability of pair production $W_{q\bar{q}}^{(1)}$
is a real quantity because $T$ is real (see Eq. (\ref{prob})).
Therefore it can be shown that the imaginary part of the above expression 
vanishes. However, for mathematical convenience we always keep the full 
expression Eq. (\ref{wqqf}) and take the real part at the end of our 
calculations. Again, this will not be denoted in the formulas, but must not be 
forgotten. 

Starting from Eq.(\ref{wqqf}) we find the source term for $q\bar{q}$
production:

\bea
\frac{dW_{q\bar{q}}}{d^4x d^3k}=
\frac{g^2}{2(2\pi)^6\omega}~A^a_{\mu}(x)
\int d^4x_2 ~A^a_{\nu}(x_2)
\int \frac{d^3\vec k_2}{2\omega_2}
\int d^4K 
~\delta^{(4)}(K-k-k_2) 
~\nonumber \\
e^{iK(x-x_2)}
{[k^{\mu} k^{\nu}_2+k^{\mu}_2k^{\nu}-\frac{K^2}{2}g^{\mu\nu}]}.
\label{dwqq0}
\eea

The $d^4K$ integration in Eq.(\ref{dwqq0}) 
can be carried out so that one finds:

\bea
\frac{dW_{q\bar{q}}^{(1)}}{d^4x d^3k}=
\frac{g^2}{2(2\pi)^6\omega}~A^a_{\mu}(x)
e^{i k \cdot x}
\int d^4x_2 ~A^a_{\nu}(x_2)~e^{-ik\cdot x_2}
\int \frac{d^3\vec k_2}{2 \omega_2} 
~\nonumber \\
~e^{i k_2 \cdot(x-x_2)}
{[k^{\mu} k^{\nu}_2+k^{\mu}_2 k^{\nu} - (m^2+k \cdot k_2)g^{\mu\nu}]}.
\label{dwqq1}
\eea

The factors of $k_2$ in Eq.(\ref{dwqq1}) can be replaced 
by differentiations with respect to $i(x_1-x_2)$ which leads to:

\bea
\frac{dW_{q\bar{q}}^{(1)}}{d^4x d^3k}=
\frac{g^2}{2(2\pi)^6\omega}~A^a_{\mu}(x)
e^{i k \cdot x}
\int d^4x_2 ~A^a_{\nu}(x_2)~e^{-ik\cdot x_2}
~\nonumber \\
{[k^{\mu} \frac{\partial}{i\partial (x-x_2)_{\nu}}
 +\frac{\partial}{i\partial (x-x_2)_{\mu}} k^{\nu}
 - (m^2+k \cdot \frac{\partial}{i\partial (x-x_2)})g^{\mu\nu}]}
\int \frac{d^3\vec k_2}{2 \omega_2}
~e^{i k_2 \cdot(x-x_2)}.
\label{dwqq2}
\eea

The $d^3\vec k_2$ integration can be performed analytically which yields

\be
\int \frac{d^3 \vec k_2}{2 \omega_2} ~e^{ik_2 \cdot(x_1-x_2)} =
4\pi m \frac{K_1(m \sqrt{-(x_1-x_2)^2})}{\sqrt{-(x_1-x_2)^2}}.
\label{bessel}
\ee

Here $K_1(z)$ stands for the modified Bessel function of the third kind and 
order one. Substituting Eq.(\ref{bessel}) into Eq.(\ref{dwqq2}) and 
performing the differentiations \footnote{For the differentiation
of the Bessel functions see {\it~e.g.} Ref. \cite{abst}} we obtain 
the general result

\bea
\frac{dW_{q\bar{q}}^{(1)}}{d^4x d^3k}=
\frac{g^2m}{(2\pi)^5\omega}~A^a_{\mu}(x)
~e^{i k \cdot x}
\int d^4x_2 ~A^a_{\nu}(x_2)~e^{-ik\cdot x_2}
~\nonumber \\
~[i(k^{\mu} (x-x_2)^{\nu}
   +(x-x_2)^{\mu} k^{\nu}
   +k \cdot (x-x_2)g^{\mu\nu})
~\nonumber \\ 
\times (\frac{K_0(m \sqrt{-(x-x_2)^2})m \sqrt{-(x-x_2)^2}+2K_1(m \sqrt{-(x-x_2)^2})}{[\sqrt{-(x-x_2)^2}]^3})
~\nonumber \\
 -m^2 g^{\mu\nu}
 \frac{K_1(m \sqrt{-(x-x_2)^2})}{\sqrt{-(x-x_2)^2}}].
\label{dwqq3}
\eea

This is the source term for the production of $q\bar{q}$ pairs from a 
space-time dependent chromofield. This source term contains all the 
information about the quark production from a space-time dependent
chromofield $A^a(x)$ in the phase space of
coordinate $x$ and momentum $k$ to the order $S^{(1)}$. 

%%%%%%%%%%%%%%%%%%%%%%%%%%%%%%%%%%%%%%%%%%%%%%%%%%%%%%%%%%%%%%%%%%%%%%%%%%%%%%%

\section{Source Term for Gluon Production from a Space-Time 
Dependent Chromofield}

For the gluons in QCD the generating functional is given by:

\be
Z[J,\xi,\xi^{\dagger}]
=\int[dA][d\chi][d\chi^{\dagger}]
\exp(iS[A,\chi,\chi^{\dagger}]
+J\cdot A+\chi\cdot\xi^{\dagger}+\chi^{\dagger}\cdot\xi).
\label{gener}
\ee

Here $J$, $\xi$, and $\xi^{\dagger}$ are external sources for the gauge field 
$A$ and the Faddeev-Popov ghosts fields $\chi^{\dagger}$ and $\chi$, 
respectively. The action $S$ is defined as follows:

\be
S=\int d^4x {\cal L}
\ee

where ${\cal L}$ consists of three terms 
${\cal L}={\cal L}_G+{\cal L}_{GF}+{\cal L}_{FP}$.
${\cal L}_G$ is the lagrangian density of the gauge field:

\be
{\cal L}_G=-\frac{1}{4}F^a_{\mu\nu}F^{a\mu\nu}
\ee

where $F^a_{\mu\nu}$ is the non-abelian field tensor defined as:
$F^a_{\mu\nu}=
\partial_{\mu}A_{\nu}-\partial_{\nu}A_{\mu}+gf^{abc}A^b_{\mu}A^c_{\nu}$

with the antisymmetric structure constant of the gauge group $f^{abc}$ and the 
coupling constant $g$.
The contribution of the gauge fixing term is given by:

\be
{\cal L}_{GF}=-\frac{1}{2\alpha}(G^a)^2
\ee

where $G^a$ is a arbitrary form linear in the $A$-field that fixes the 
gauge. For the common choice 
$G^a=\partial^{\mu}A^a_{\mu}$
the Faddeev-Popov ghost term becomes:
${\cal L}_{FP}=-(\partial_{\mu}\chi^{\dagger}_a)(D^{\mu}\chi)_a$ where 
$D^{\mu}$ denotes the covariant derivative 
$D^{\mu}=\partial^{\mu}+gT^aA^{a\mu}$ with the generators of the gauge group 
in adjoint representation $T^a$.
The scalar products in Eq. (\ref{gener}) are defined as:

\be
J\cdot A=\int d^4x J_{\mu}A^{\mu}.
\ee

In the background field method of QCD, the gauge field is split into two parts:

\be
A_{\mu}\rightarrow A_{\mu}+Q_{\mu}.
\ee

From here on, $A_{\mu}$ is a classical, {\it i.e.} non-quantized background 
field and $Q_{\mu}$ is a quantum field representing the gluons. The lagrangian 
density of the gauge field now becomes a functional of $A$ and $Q$:

\be
{\cal L}_G=-\frac{1}{4}F^a_{\mu\nu}[A+Q]F^{a\mu\nu}[A+Q].
\ee

Additionally, a special gauge is chosen. Only a gauge for $Q$ has got to be 
fixed, as only this field is going to be quantized. The so called background 
field gauge is given by \cite{thooft}:

\be
G^a=
\partial^{\mu}Q^a_{\mu}+gf^{abc}A^{b\mu}Q^c_{\mu}=
(D^{\mu}[A]Q_{\mu}^a).
\label{gauge}
\ee

Due to this choice, the ghost term becomes:

\be
{\cal L}_{FP}=-(D_{\mu}[A]\chi^{\dagger}_a)(D^{\mu}[A+Q]\chi^a).
\ee

The total lagrangian density $\cal L$ in the background field formalism is 
gauge invariant with respect 
to the type I gauge transformations \cite{izju}:

\be
A_{\mu}\rightarrow A'_{\mu}=UA_{\mu}U^{-1}-\frac{i}{g}(\partial_{\mu}U)U^{-1},
\label{trafo1}
\ee 

\be
Q_{\mu}\rightarrow Q'_{\mu}=UQ_{\mu}U^{-1},
\label{trafo2}
\ee

and

\be
\chi \rightarrow \chi{'}=U \chi U^{-1},
\label{trafo3}
\ee

even after quantization. In the following, all calculations are to be performed
in the background field Feynman gauge, {\it i.e.} with $\alpha=1$.

For the gluon pair production from a space-time dependent chromofield in 
the order of $S^{(1)}$ the vacuum-polarization amplitude is given by:

\be
M=<k_1k_2|S^{(1)}|0>.
\ee

To this order those terms of the interaction lagrangian 
${\cal L}_I$ involving two $Q$-fields are given by: 

\bea
{\cal L}_I^{(1)}=
-\frac{1}{2}F^a_{\mu\nu}[A]gf^{abc}Q^{b\mu}Q^{c\nu}
-\frac{1}{2}(\partial_{\mu}Q^a_{\nu}-\partial_{\nu}Q^a_{\mu})
    gf^{abc}(A^{b\mu}Q^{c\nu}+Q^{b\mu}A^{c\nu})
~\nonumber \\
-\frac{1}{4}g^2f^{abc}f^{ab'c'}(A^b_{\mu}Q^c_{\nu}+Q^b_{\mu}A^c_{\nu})
                               (A^{b'\mu}Q^{c'\nu}+Q^{b'\mu}A^{c'\nu})
~\nonumber \\
-\partial_{\lambda}Q^{a\lambda}gf^{abc}A^b_{\kappa}Q^{c\kappa}
-\frac{1}{2}g^2f^{abc}f^{ab'c'}A^b_{\lambda}Q^{c\lambda}
                               A^{b'}_{\kappa}Q^{c'\kappa}.
\label{tot}
\eea

From these terms one can read off the necessary Feynman rules. For the 
production of two gluons by coupling to the field once 
(see Fig.(2a)), one obtains the vertex: 

\be
(V_{1A})^{abd}_{\mu\nu\rho}=
gf^{abd}[-2g_{\mu\rho}K_{\nu}
-g_{\nu\rho}(k_1-k_2)_{\mu}
+2g_{\mu\nu}K_{\rho}],
\label{V1A}
\ee

where $K=k_1+k_2$.
If two couplings to the field are involved (see Fig.(2b)), 
one receives: 

\bea
(V_{2A})^{abcd}_{\mu\nu\lambda\rho}=-ig^2
[f^{abx}f^{xcd}
(g_{\mu\lambda}g_{\nu\rho}-g_{\mu\rho}g_{\nu\lambda}+g_{\mu\nu}g_{\lambda\rho})
~\nonumber \\
+f^{adx}f^{xbc}
(g_{\mu\nu}g_{\lambda\rho}-g_{\mu\lambda}g_{\nu\rho}-g_{\mu\rho}g_{\nu\lambda})
~\nonumber \\
+f^{acx}f^{xbd}
(g_{\mu\nu}g_{\lambda\rho}-g_{\mu\rho}g_{\nu\lambda})].
\label{V2A}
\eea

The above rules coincide with those derived in \cite{abbott}. The total 
amplitude for the production of gluon pairs is now equal to the sum of the 
single contributions:

\be
M=M_{1A}+M_{2A}.
\ee

If the correct weight factors are included, so that the lagrangian density can 
again be retrieved, one finds for the first term:

\bea
M_{1A}=
\frac{(2\pi)^2}{2}\int d^4K\delta^{(4)}(K-k_1-k_2)
A^{a\mu}(K)\epsilon^{b\nu}(k_1)\epsilon^{d\rho}(k_2)(V_{1A})^{abd}_{\mu\nu\rho}
\label{ampy}
\eea

where finally the allowed kinematical region of integration is given by 
$(K)^2>0$ and $K^0>0$. The second term is given by:

\bea
M_{2A}=
\frac{1}{4}\int d^4k_3 d^4k_4\delta^{(4)}(k_1+k_2-k_3-k_4)
%~\nonumber \\
A^{a\mu}(k_3)A^{c\lambda}(k_4)\epsilon^{b\nu}(k_1)\epsilon^{d\rho}(k_2)
(V_{2A})^{abcd}_{\mu\nu\lambda\rho},
\label{ampx}
\eea

where the limitations $(k_3+k_4)^2>0$ and $k_3^0+k_4^0>0$ have to be obeyed
in the end. 
The probability can again be obtained by making use of Eq.(\ref{prob}). In 
order to obtain the physical polarization of the gluons, we use the following 
spin-sums:

\be
\sum_{spin}\epsilon^{\nu}(k_1)\epsilon^{*\nu'}(k_1)=
\sum_{spin}\epsilon^{\nu}(k_2)\epsilon^{*\nu'}(k_2)=-g^{\nu\nu'},
\ee

but later on deduct the corresponding probabilities for the processes 
involving ghost instead of gluons as shown in Fig.(3). In our 
case, the total 
probability for the production of gluons not yet corrected for ghosts 
consists of four terms which add up:

\be
W^A=W_{1A,1A}+W_{1A,2A}+W_{2A,1A}+W_{2A,2A}.
\label{wa}
\ee

By definition, we have for the first term in Eq. (\ref{wa}):

\be
W_{1A,1A}
=\frac{1}{2}
\sum_{spin}\int\frac{d^3k_1}{(2\pi)^32k_1^0}\frac{d^3k_2}{(2\pi)^32k_2^0}
M_{1A}M^*_{1A}.
\label{D1A1A}
\ee

For the integration over the phase-space of the gluons there is an additional 
factor of $1/2$ because the particles in the final state are identical.
The introduction of the Fourier transform Eq.(\ref{fourier}) and the 
application of

\be
f^{abd}f^{a'bd}=3\delta^{aa'}
\label{ff}
\ee

yields:

\bea
W_{1A,1A}
=\frac{3g^2}{8(2\pi)^6}
\int\frac{d^3k_1}{k_1^0}\frac{d^3k_2}{k_2^0}d^4xd^4x'
e^{i(k_1+k_2)\cdot(x-x')}
A^{a\mu}(x)A^{a\mu'}(x')
~\nonumber \\
~[2g_{\mu\mu'}(k_1+k_2)^2
 -2(k_1+k_2)_{\mu}(k_1+k_2)_{\mu'}
 +(k_1-k_2)_{\mu}(k_1-k_2)_{\mu'}].
\label{W1A1A}
\eea

Again, it has to be noted that the real part of this quantity has to be taken. 
By the same means as in the quark case, we obtain the first contribution to 
the gluon source term:

\bea
\frac{dW_{1A,1A}}{d^4xd^3k}
=\frac{3g^2}{8(2\pi)^6}
\int\frac{d^3k_2}{k^0k_2^0}d^4x'
e^{i(k+k_2)\cdot(x-x')}
A^{a\mu}(x)A^{a\mu'}(x')
~\nonumber \\
~[2g_{\mu\mu'}(k+k_2)^2
 -2(k+k_2)_{\mu}(k+k_2)_{\mu'}
 +(k-k_2)_{\mu}(k-k_2)_{\mu'}].
\label{dW1A1A_}
\eea

Making use of the following relation

\be
\int\frac{d^3k_2}{|\vec k_2|}e^{ik_2\cdot(x-x')}=-\frac{8\pi}{(x-x')^2},
\label{int1}
\ee

and explicitly calculating the derivatives afterwards results in the final 
form (see Appendix):

\bea
\frac{dW_{1A,1A}}{d^4xd^3k}
=\frac{3g^2}{2(2\pi)^5 k^0}
\int d^4x'
e^{ik\cdot(x-x')}
A^{a\mu}(x)A^{a\mu'}(x')
[k_{\mu}k_{\mu'}
%~\nonumber \\
-4g_{\mu\mu'}k^{\nu}i\frac{(x-x')_{\nu}}{(x-x')^2}
~\nonumber \\
+3(k_{\mu}i\frac{(x-x')_{\mu'}}{(x-x')^2}
  +k_{\mu'}i\frac{(x-x')_{\mu}}{(x-x')^2})
+2\frac{g_{\mu\mu'}}{(x-x')^2}-4\frac{(x-x')_{\mu}(x-x')_{\mu'}}{(x-x')^4}]
\frac{1}{(x-x')^2}.
\label{dW1A1A}
\eea

Now we perform the same procedure with the other contributions. First for the 
two mixed contributions we get:

\bea
W_{1A,2A}=W_{2A,1A}=
\frac{1}{2}
\sum_{spin}\int\frac{d^3k_1}{(2\pi)^32k_1^0}\frac{d^3k_2}{(2\pi)^32k_2^0}
M_{2A}M^*_{1A}.
\label{W1A2A}
\eea

After expressing the probability in terms of the Fourier transforms of the 
$A$-fields:

\bea
W_{1A,2A}=W_{2A,1A}=
~\nonumber \\
\frac{-ig^3}{64(2\pi)^{10}}
\int\frac{d^3k_1}{k_1^0}\frac{d^3k_2}{k_2^0}d^4k_3 d^4k_4d^4Kd^4x_3d^4x_4d^4x
\delta^{(4)}(K-k_3-k_4)\delta^{(4)}(K-k_1-k_2)
~\nonumber \\
e^{i[k_3\cdot x_3+k_4\cdot x_4-K\cdot x]}
A^{a\mu}(x_3)A^{c\lambda}(x_4)A^{a'\mu'}(x)
24f^{a'ac}K_{\lambda}g_{\mu\mu'},
\eea

we can extract the source term:

\bea
\frac{dW_{2A,1A}}{d^4xd^3k}
=\frac{-3ig^3}{8(2\pi)^{10}}
\int\frac{d^3k_2}{k^0k_2^0}d^4k_3 d^4k_4d^4Kd^4x_3d^4x_4
\delta^{(4)}(K-k_3-k_4)\delta^{(4)}(K-k-k_2)
~\nonumber \\
e^{i[k_3\cdot x_3+k_4\cdot x_4-K\cdot x]}
A^{a\mu}(x_3)A^{c\lambda}(x_4)A^{a'\mu'}(x)
f^{a'ac}K_{\lambda}g_{\mu\mu'}.
\label{dW2A1A_}
\eea

At this point, we again need Eq. (\ref{int1}) to find (see Appendix):

\bea
\frac{dW_{2A,1A}}{d^4xd^3k}
=\frac{3ig^3}{2(2\pi)^5k^0}
\int d^4x'
e^{ik\cdot (x'-x)}
A^{c\lambda}(x')
(A^{a}(x')\cdot A^{a'}(x))
f^{a'ac}(\frac{k_{\lambda}}{(x'-x)^2}+2i\frac{(x'-x)_{\lambda}}{(x'-x)^4}).
\label{dW2A1A}
\eea

And finally for the last term: From the definition

\bea
W_{2A,2A}
=\frac{1}{2}
\sum_{spin}\int\frac{d^3k_1}{(2\pi)^32k_1^0}\frac{d^3k_2}{(2\pi)^32k_2^0}
M_{2A}M^*_{2A}
\eea

we obtain

\bea
W_{2A,2A}
=\frac{g^4}{64(2\pi)^{14}}
\int\frac{d^3k_1}{k_1^0}\frac{d^3k_2}{k_2^0}
d^4k_3 d^4k_4d^4k'_3 d^4k'_4d^4x_3d^4x_4d^4x'_3d^4x'_4
e^{i[k_3\cdot x_3+k_4\cdot x_4-k'_3\cdot x'_3-k'_4\cdot x'_4]}
~\nonumber \\
\delta^{(4)}(k_1+k_2-k_3-k_4)\delta^{(4)}(k_3+k_4-k'_3-k'_4)
A^{a\mu}(x_3)A^{c\lambda}(x_4)A^{a'\mu'}(x'_3)A^{c'\lambda'}(x'_4)
~\nonumber \\
~[2g_{\mu\lambda}g_{\mu'\lambda'}(f^{abx}f^{xcd}+f^{adx}f^{xcb})
                                 (f^{a'bx'}f^{x'c'd}+f^{a'dx'}f^{x'c'b})
 +24g_{\mu\mu'}g_{\lambda\lambda'}f^{acx}f^{a'c'x}].
\label{W2A2A}
\eea

And from there the source term:

\bea
\frac{dW_{2A,2A}}{d^4xd^3k}
=\frac{g^4}{32(2\pi)^{14}}
\int\frac{d^3k_2}{k^0k_2^0}
d^4k_3 d^4k_4d^4k'_3 d^4k'_4d^4x_4d^4x'_3d^4x'_4
e^{i[k_3\cdot x+k_4\cdot x_4-k'_3\cdot x'_3-k'_4\cdot x'_4]}
~\nonumber \\
\delta^{(4)}(k+k_2-k_3-k_4)\delta^{(4)}(k_3+k_4-k'_3-k'_4)
A^{a\mu}(x)A^{c\lambda}(x_4)A^{a'\mu'}(x'_3)A^{c'\lambda'}(x'_4)
~\nonumber \\
~[g_{\mu\lambda}g_{\mu'\lambda'}(f^{abx}f^{xcd}+f^{adx}f^{xcb})
                                 (f^{a'bx'}f^{x'c'd}+f^{a'dx'}f^{x'c'b})
 +12g_{\mu\mu'}g_{\lambda\lambda'}f^{acx}f^{a'c'x}].
\label{dW2A2A_}
\eea

Here we apply Eq. (\ref{int1}) to receive (see Appendix)

\bea
\frac{dW_{2A,2A}}{d^4xd^3k}
=\frac{-g^4}{8(2\pi)^5k^0}
\int d^4x'_3
e^{ik\cdot(x-x'_3)}
\frac{1}{(x-x'_3)^2}
A^{a\mu}(x)A^{c\lambda}(x)A^{a'\mu'}(x'_3)A^{c'\lambda'}(x'_3)
~\nonumber \\
~[g_{\mu\lambda}g_{\mu'\lambda'}(f^{abx}f^{xcd}+f^{adx}f^{xcb})
                                 (f^{a'bx'}f^{x'c'd}+f^{a'dx'}f^{x'c'b})
 +12g_{\mu\mu'}g_{\lambda\lambda'}f^{acx}f^{a'c'x}].
\label{dW2A2A}
\eea

As already remarked before, we also have to calculate the corresponding terms 
for the 
ghosts in order to be able to remove the contributions by unphysical 
polarizations of the gluons. To the ghost matrix element, there contribute the 
two Feynman 
diagrams shown in Fig. (3) with the interaction vertices

\be
(V^{FP}_{1A})^{abd}_{\mu}=+gf^{abd}(k_1-k_2)_{\mu},
\label{VFP1A}
\ee

and

\be
(V^{FP}_{2A})^{abcd}_{\mu\lambda}=
-ig^2g_{\mu\lambda}(f^{abx}f^{xcd}+f^{adx}f^{xcb}).
\label{VFP2A}
\ee

These vertices are obtained from the ghost lagrangian density:

\be
{\cal L}_{FP}^{(I)}=
- (\partial_{\mu}\chi^{a\dagger})f^{abc}A_{\mu}^b\chi^c
- (f^{abc}A_{\mu}^b[A]\chi^{c\dagger})\partial_{\mu}\chi^a
- (f^{abc}A_{\mu}^b\chi^{c\dagger})f^{ade}A_{\mu}^d\chi^e
\label{lfpi}
\ee

Including the correct weight factors we obtain the amplitude:

\be
(M^{FP})^{bd}=(M^{FP}_{1A})^{bd}+(M^{FP}_{2A})^{bd}
\ee

with

\bea
(M_{1A}^{FP})^{bd}=
\frac{(2\pi)^2}{2}\int d^4K\delta^{(4)}(k_1+k_2-K)
A^{a\mu}(K)(V^{FP}_{1A})^{abd}_{\mu}
\label{MFP1A}
\eea

and

\bea
(M_{2A}^{FP})^{bd}=
\frac{1}{4}\int d^4k_3 d^4k_4\delta^{(4)}(k_1+k_2-k_3-k_4)
A^{a\mu}(k_3)A^{c\lambda}(k_4)(V^{FP}_{2A})^{abcd}_{\mu\lambda}.
\label{MFP2A}
\eea

The probability is defined as for the gluons but without the factor $1/2$ for 
the phase space:

\be
W^{FP}=
\int\frac{d^3k_1}{(2\pi)^32k_1^0}\frac{d^3k_2}{(2\pi)^32k_2^0}
(M^{FP})^{bd}(M^{FP})^{*bd}.
\ee

This time, there are only two contributions, because the cross-terms vanish:

\be
W^{FP}=W^{FP}_{1A,1A}+W^{FP}_{2A,2A}.
\ee

Starting from the definition of the first term

\bea
W_{1A,1A}^{FP}=
\int\frac{d^3k_1}{(2\pi)^32k_1^0}\frac{d^3k_2}{(2\pi)^32k_2^0}
(M^{FP}_{1A})^{bd}(M^{FP}_{1A})^{*bd},
\label{W1A1AFP}
\eea

we carry out the same steps as for the derivation of Eq. (\ref{dW1A1A}) from 
Eq. (\ref{D1A1A}) through Eq.(\ref{W1A1A}). From 

\bea
W_{1A,1A}^{FP}
=\frac{3g^2}{16(2\pi)^2}
\int\frac{d^3k_1}{k_1^0}\frac{d^3k_2}{k_2^0}d^4K
\delta^{(4)}(k_1+k_2-K)
A^{a\mu}(K)A^{*a\mu'}(K)
(k_1-k_2)_{\mu}(k_1-k_2)_{\mu'}
\eea

we extract the source term:

\bea
\frac{dW^{FP}_{1A,1A}}{d^4xd^3k}=
\frac{3g^2}{4(2\pi)^5k^0}
\int d^4x'
e^{ik\cdot(x-x')}
A^{a\mu}(x)A^{a\mu'}(x')
~\nonumber \\
~[k_{\mu}i\frac{(x-x')_{\mu'}}{(x-x')^2}
+k_{\mu'}i\frac{(x-x')_{\mu}}{(x-x')^2}
-k_{\mu}k_{\mu'}-
2\frac{g_{\mu\mu'}}{(x-x')^2}+4\frac{(x-x')_{\mu}(x-x')_{\mu'}}{(x-x')^4}]
\frac{1}{(x-x')^2}.
\label{dWFP1A1A}
\eea

Analogously, we obtain the second contribution from its definition:

\bea
W_{2A,2A}^{FP}=
\int\frac{d^3k_1}{(2\pi)^32k_1^0}\frac{d^3k_2}{(2\pi)^32k_2^0}
(M^{FP}_{2A})^{bd}(M^{FP}_{2A})^{*bd}
\label{W2A2AFP}
\eea

and finally get:

\bea
W_{2A,2A}^{FP}
=\frac{g^4}{64(2\pi)^6}
\int\frac{d^3k_1}{k_1^0}\frac{d^3k_2}{k_2^0}d^4k_3 d^4k_4d^4k'_3 d^4k'_4
\delta^{(4)}(k_1+k_2-k_3-k_4)\delta^{(4)}(k_1+k_2-k'_3-k'_4)
~\nonumber \\
A^{a\mu}(k_3)A^{c\lambda}(k_4)A^{*a'\mu'}(k'_3)A^{*c'\lambda'}(k'_4)
g_{\mu\lambda}g_{\mu'\lambda'}
(f^{abx}f^{xcd}+f^{adx}f^{xcb})(f^{a'bx'}f^{x'c'd}+f^{a'dx'}f^{x'c'b}).
\eea

Comparison to Eq.(\ref{dW2A2A}) yields the source term:

\bea
\frac{dW_{2A,2A}^{FP}}{d^4xd^3k}
=\frac{-g^4}{16(2\pi)^5k^0}
\int d^4x'_3
e^{ik\cdot(x-x'_3)}
\frac{1}{(x-x'_3)^2}
A^{a\mu}(x)A^{c\lambda}(x)A^{a'\mu'}(x'_3)A^{c'\lambda'}(x'_3)
~\nonumber \\
g_{\mu\lambda}g_{\mu'\lambda'}
(f^{abx}f^{xcd}+f^{adx}f^{xbc})(f^{a'bx'}f^{x'c'd}+f^{a'dx'}f^{x'bc'}).
\label{dWFP2A2A}
\eea

In the end, we obtain the required source term for the production of gluon 
pairs from a space-time dependent chromofield via vacuum-polarization by 
combining all the contributions in the following manner:

\be
W_{gg}=W_{1A,1A}-W_{1A,1A}^{FP}+2W_{2A,1A}+W_{2A,2A}-W_{2A,2A}^{FP}. 
\ee

So, we receive:

\bea
\frac{dW_{gg}}{d^4xd^3k}
=
\frac{1}{(2\pi)^5 k^0}\int d^4x' e^{ik\cdot(x-x')}\frac{1}{(x-x')^2}\{
\frac{3}{4}g^2
A^{a\mu}(x)A^{a\mu'}(x')
[3k_{\mu}k_{\mu'}
-8g_{\mu\mu'}k^{\nu}i\frac{(x-x')_{\nu}}{(x-x')^2}
~\nonumber \\
+5(k_{\mu}i\frac{(x-x')_{\mu'}}{(x-x')^2}
  +k_{\mu'}i\frac{(x-x')_{\mu}}{(x-x')^2})
+6\frac{g_{\mu\mu'}}{(x-x')^2}-12\frac{(x-x')_{\mu}(x-x')_{\mu'}}{(x-x')^4}]
~\nonumber \\
-
3ig^3
A^{a\mu}(x')A^{c\lambda}(x')A^{a'\mu'}(x)
f^{a'ac}K_{\lambda}g_{\mu\mu'}
~\nonumber \\
-
\frac{1}{16}g^4
A^{a\mu}(x)A^{c\lambda}(x)A^{a'\mu'}(x')A^{c'\lambda'}(x')
~\nonumber \\
~[g_{\mu\lambda}g_{\mu'\lambda'}(f^{abx}f^{xcd}+f^{adx}f^{xcb})
                                 (f^{a'bx'}f^{x'c'd}+f^{a'dx'}f^{x'c'b})
 +24g_{\mu\mu'}g_{\lambda\lambda'}f^{acx}f^{a'c'x}]
\}.
\label{dwgg}
\eea

This is the source term for the production of gluon pairs for any arbitrary 
space-time dependent chromofield in the order $S^{(1)}$. 

It can be checked that this source term is gauge invariant with respect to 
type-(I)-gauge transformations in the following manner (see also \cite{dng}). 
One can observe that
the gauge invariant part of the lagrangian which in each term contains two $Q$ 
fields or two ghost fields is given by:

\bea
{\cal L} = -\frac{1}{2} gf^{abc}~F^a_{\mu\nu}[A]Q^{b\mu}Q^{c\nu} \nonumber \\
{-\frac{1}{4} [D_{\mu}[A]Q^{a\nu}-D_{\nu}[A]Q^{a\mu}]
[D_{\mu}[A]Q^{a\nu}-D_{\nu}[A]Q^{a\mu}]} \nonumber \\
{-\frac{1}{2\alpha}
[D_{\mu}[A]Q^{a\mu}]^2 - (D_{\mu}[A]\chi^{\dagger}_a)D_{\mu}[A]\chi^a}.
\label{tott}
\eea

This lagrangian density is gauge invariant with respect to the gauge 
transformations in Eqs.(\ref{trafo1}), (\ref{trafo2}), and (\ref{trafo3}).
It can be checked that the free terms, {\it i.e.} terms not involving $A$, of 
the gauge invariant
lagrangian density (Eq. (\ref{tott})), when evaluated between an initial
vacuum state $|0>$ and a final physical two gluon state $<k_1,k_2|$, do not
contribute to the probability. Hence, the expression we have used
for $\cal L$ in Eq. (\ref{tot}) together with Eq. (\ref{lfpi}) for the ghosts 
gives the correct result
for the gluon pair production probability, invariant with respect to the above 
gauge transformations.

%%%%%%%%%%%%%%%%%%%%%%%%%%%%%%%%%%%%%%%%%%%%%%%%%%%%%%%%%%%%%%%%%%%%%%%%%%%%%%%

\section{Source Term for Parton Production from a Purely Time Dependent 
Chromofield}

Our main purpose is to use the obtained source terms of quarks and gluons in 
the relativistic non-abelian transport equations to study the production and
equilibration of the quark-gluon plasma expected to be formed
in ultra relativistic heavy-ion collisions (URHIC). Solving the 
transport equations with all the effects,
such as collisions among the partons, acceleration of the partons by the
background field, precession of the color charge in the group space
and production of the partons from the background field involves 
much more numerical work as was done in \cite{num}. In the following
we are going to analyze an example: We will consider a time dependent
chromofield and examine the behavior of the source term in the
phase-space. Solutions of the transport equations including these source terms 
to study the production and equilibration of quark-gluon plasma will be 
presented elsewhere.

% % % % % % % % % % % % % % % % % % % % % % % % % % % % % % % % % % % % % % % %

\subsection{Source Terms for any Arbitrary Time Dependent Chromofield}

For a purely time-dependent field Eq. (\ref{dwqq0}) reduces to:

\bea
\frac{dW_{q\bar{q}}}{d^4x d^3k}=
\frac{g^2}{2(2\pi)^3\omega}~A^a_{\mu}(t)
\int dt_2 ~A^a_{\nu}(t_2)
\int \frac{d^3\vec k_2}{2 \omega_2}
\int d^4K 
~\delta^{(4)}(K-k-k_2) ~\delta^{(3)}(\vec K) 
~\nonumber \\
~e^{i(K\cdot x-K^0t_2)}
{[k^{\mu} k^{\nu}_2+k^{\mu}_2 k^{\nu} - \frac{K^2}{2}g^{\mu\nu}]}.
\label{dwqqt0}
\eea

After integrating over $d^4K$ and $d^3 \vec k_2$ and taking care of the 
Dirac $\delta$ factors we find

\bea
\frac{dW_{q\bar{q}}}{d^4x d^3k}=
\frac{g^2}{2(2\pi)^3\omega}
\int dt_2
%~\nonumber \\
\frac{e^{2i\omega(t-t_2)}}{2\omega}
[2\omega^2(\vec A^a(t)\cdot\vec A^a(t_2))
-2(\vec A^a(t)\cdot\vec k)(\vec A^a(t_2)\cdot\vec k)].
\label{dwqqt0a}
\eea

Here, we have presented Eq. (\ref{dwqqt0a}) in the form of scalar products of 
three-vectors.
The last remaining time integration then constitutes a Fourier 
transform from $t_2$ to $2\omega_1$. The final result for the source 
term in the case of a time-dependent chromofield is 

\bea
\frac{dW_{q\bar{q}}}{d^4x d^3k}=
\frac{g^2\sqrt{2\pi}}{2(2\pi)^3\omega^2}
e^{2i\omega t}
[ \omega^2(\vec A^a(t) \cdot \vec A^{*a}(2 \omega))
 -(\vec A^a(t) \cdot \vec k) (\vec A^{*a}(2 \omega) \cdot \vec k)].
\label{dwqqt1}
\eea

%  %  %  %  %  %  %  %  %  %  %  %  %  %  %  %  %  %  %  %  %  %  %  %  %  %  %

Analogously, the Eqs. (\ref{dW1A1A_},\ref{dW2A1A_}, and \ref{dW2A2A_}) 
for the gluons and the Eqs. (\ref{dWFP1A1A} and \ref{dWFP2A2A}) for the ghosts 
simplify to (see Appendix): 

\bea
\frac{dW_{1A,1A}}{d^4xd^3k}
=\frac{3g^2\sqrt{2\pi}}{2(2\pi)^3}
e^{2ik^0t}
~[(\vec A^a(t)\cdot\frac{\vec k}{k^0})
  (\vec A^{*a}(2k^0)\cdot\frac{\vec k}{k^0})
  -2(\vec A^{a}(t)\cdot\vec A^{*a}(2k^0))],
\label{dW1A1At}
\eea

\bea
\frac{dW_{2A,1A}}{d^4xd^3k}
=\frac{-3ig^3}{4(2\pi)^3k^0}
\int dt_3
e^{2ik^0(t_3-t)}
A^{a\mu}(t_3)A^{c0}(t_3)A^{a'\mu'}(t)
f^{a'ac}g_{\mu\mu'},
\label{dW2A1At}\eea

\bea
\frac{dW_{2A,2A}}{d^4xd^3k}
=\frac{g^4}{32(2\pi)^3(k^0)^2}
\int dt'_3
e^{2ik^0(t-t'_3)}
A^{a\mu}(t)A^{c\lambda}(t)A^{a'\mu'}(t'_3)A^{c'\lambda'}(t'_3)
~\nonumber \\
~[g_{\mu\lambda}g_{\mu'\lambda'}(f^{abx}f^{xcd}+f^{adx}f^{xcb})
                                 (f^{a'bx'}f^{x'c'd}+f^{a'dx'}f^{x'c'b})
 +12g_{\mu\mu'}g_{\lambda\lambda'}f^{acx}f^{a'c'x}],
\label{dW2A2At}
\eea

\bea
\frac{dW_{1A,1A}^{FP}}{d^4xd^3k}
=\frac{3g^2\sqrt{2\pi}}{4(2\pi)^3}
e^{2ik^0t}
(\vec A^{a}(t)\cdot\frac{\vec k}{k^0})
(\vec A^{*a}(2k^0)\cdot\frac{\vec k}{k^0}),
\eea

and

\bea
\frac{dW_{2A,2A}^{FP}}{d^4xd^3k}
=\frac{g^4}{64(2\pi)^3(k^0)^2}
\int
dt'_3
e^{2ik^0(t-t'_3)}
A^{a\mu}(t)A^{c\lambda}(t)A^{a'\mu'}(t'_3)A^{c'\lambda'}(t'_3)
~\nonumber \\
g_{\mu\lambda}g_{\mu'\lambda'}
(f^{abx}f^{xcd}+f^{adx}f^{xcb})(f^{a'bx'}f^{x'c'd}+f^{a'dx'}f^{x'c'b}).
\eea

Adding the gluon and deducting the ghost contributions, we obtain:

\bea
\frac{dW_{gg}}{d^4xd^3k}
=
e^{2ik^0t}\frac{\sqrt{2\pi}}{(2\pi)^3}
\{
\frac{3g^2}{4}
~[(\vec A^a(t)\cdot\frac{\vec k}{k^0})
  (\vec A^{*a}(2k^0)\cdot\frac{\vec k}{k^0})
  -4(\vec A^{a}(t)\cdot\vec A^{*a}(2k^0))]
~\nonumber \\
+\frac{6ig^3}{4k^0}
(A^{a\mu}*A^{c0})^*(2k^0)A^{a'\mu'}(t)
f^{a'ac}g_{\mu\mu'}
~\nonumber \\
+\frac{g^4}{64(k^0)^2} 
A^{a\mu}(t)A^{c\lambda}(t)(A^{a'\mu'}*A^{c'\lambda'})^*(2k^0)
~\nonumber \\
~[g_{\mu\lambda}g_{\mu'\lambda'}(f^{abx}f^{xcd}+f^{adx}f^{xcb})
                                 (f^{a'bx'}f^{x'c'd}+f^{a'dx'}f^{x'c'b})
 +24g_{\mu\mu'}g_{\lambda\lambda'}f^{acx}f^{a'c'x}]\}.
\label{dWtott}
\eea

This is the general source term for the production of gluon pairs in the 
presence of a purely time dependent chromofield.

% % % % % % % % % % % % % % % % % % % % % % % % % % % % % % % % % % % % % % % %

\subsection{A Special Case}

We now choose a special form of the field in order to get some insight into
the behavior of the obtained source terms for quarks and gluons. 
For simplicity we choose the field to be

\be
A^{a3}(t)=A_{in}e^{-|t|/t_0},~t_0>0,~a=1,...,8
\label{as},
\ee

and all other components are equal to zero. 
Many other forms could have been taken. We have 
chosen this option just to get a feeling for how the source term in the 
phase-space behaves. This choice is also inspired from the numerical study 
\cite{num} which shows that the decay of the field is close to this behavior.
In any case the actual form of the decay of the classical field can only be 
determined from a self consistent solution of the relativistic non-abelian 
transport equations, as the transport equations have to be combined with the 
energy momentum conservation equations for particles and fields. That is why 
the exact form for the decay of the fields (due to the production, 
acceleration of partons, precession of color, collision among partons, and 
expansion of the system) can be determined from self consistent
transport studies (see \cite{num} for details). We will use the general source 
terms (see section III and IV) in the relativistic transport equations to 
describe the production and equilibration of the quark-gluon plasma in 
ultra relativistic heavy-ion collisions in the future.

The Fourier transform of Eq. (\ref{as}) is given by

\be
A^{a3}(-2|\vec k|)=
A_{in}\frac{2}{\sqrt{2\pi}}\frac{t_0}{1+4|\vec k|^2t_0^2}.
\label{asf}
\ee

Putting Eqs. (\ref{as}) and (\ref{asf}) into Eq.(\ref{dwqqt1}) and afterwards 
summing over color space yields:

\bea
\frac{dW_{q\bar{q}}}{d^4x d^3k}=
16\frac{\alpha_S}{(2\pi)^2}
(A_{in})^2
e^{2 i \omega t}
e^{-|t|/t_0}
\frac{t_0}{1+4\omega^2t_0^2}
\frac{m_T^2}{\omega^2},
\label{dWqqs}
\eea

with $m_T^2=m^2+k_T^2$ and where $k_T$ is the transverse momentum.
Using, in the same way, Eqs. (\ref{as}) and (\ref{asf}) in Eq.(\ref{dWtott}) 
results in:

\bea
\frac{dW_{gg}}{d^4xd^3k}
=
\frac{24\alpha_S}{(2\pi)^2}(A_{in})^2e^{2ik^0t}e^{-|t|/t_0}
\frac{t_0}{1+4(k^0)^2t_0^2}(-3-\frac{k_T^2}{(k^0)^2})
~\nonumber \\
+
\frac{36\alpha_S^2}{2\pi}(A_{in})^4e^{2ik^0t}e^{-2|t|/t_0}
\frac{t_0}{1+(k^0)^2t_0^2}\frac{1}{(k^0)^2}.
\label{dWtots}
\eea

The contribution of the mixed term (\ref{dW2A1A}) vanishes for all fields of 
the form $A^{a\mu}(x)=A^{a\mu}_{in}f(x)$. For our choice the 
contribution of the last term in Eq. (\ref{dW2A2A}) is zero.

In the next chapter, we will present plots for the source terms derived above 
(Eqs. (\ref{dWqqs}) and (\ref{dWtots})) for our choice of input parameters. 

%%%%%%%%%%%%%%%%%%%%%%%%%%%%%%%%%%%%%%%%%%%%%%%%%%%%%%%%%%%%%%%%%%%%%%%%%%%%%%%

\section{Results and Discussion}

The general results for the source terms for the production of quark-antiquark 
pairs and gluon pairs by vacuum polarization in the presence of a general 
classical space-time dependent chromofield to the order $S^{(1)}$ are given 
by Eq.(\ref{dwqq3}) and Eq.(\ref{dwgg}), respectively. 
Continuing from this point, {\it i.e.} using these source terms in relativistic 
non-abelian transport equations to study the production and equilibration of 
the quark-gluon plasma, involves extensive 
numerical computations \cite{num} which have to be represented later on.
In order to get an idea about the behavior of the source terms, we 
derived the general expressions for them in the presence of a 
purely time-dependent field. Subsequently, we assumed a certain, physically 
motivated form for the background field, given in Eq. (\ref{as}) to obtain 
Eqs. (\ref{dWqqs}) and (\ref{dWtots}). We now present the results from the 
above equations.

We relate the energy $k_0$ of the produced particles to the rapidity $y$ by:

\be
k^0=k_T\cosh(y),
\ee

where $k_T$ is the transverse momentum. 
If not stated otherwise, we choose the following parameters: $\alpha_S=0.15$, 
$A_{in}=1.5GeV$, $k_T=1.5GeV$, $y=0$, and $t_0=0.5fm$ and the quarks are 
considered to be massless. These values might not correspond to an exact 
combination of values in an URHIC. We have just chosen a set of values as an 
example in order to demonstrate the properties of our source term.

In Fig. (4), we plot the source term

\be
S=\frac{dW}{d^4xd^3k}.
\ee

for quarks (Eq. (\ref{dWqqs})) and for gluons (Eq. (\ref{dWtots})) versus 
time for the above choice of parameters. All quark graphs are to be multiplied 
by a factor of two.
The exponential decay originates mostly from the decay of the model-field. 
Of course, every decaying field will transfer this quality to the production 
rate, but this is not necessarily an exponential decrease. 
The oscillatory behavior is due to the exponential factor with imaginary 
exponent. This factor is already present in the general formula. 
Physically, it seems to indicate that there exist periods of particle creation 
and particle annihilation which follow each other periodically.
Additionally, the frequency of this oscillation is 
varying with the energy of the produced particles.
This oscillatory behavior of the source term will play a crucial role once 
it is included in a self consistent transport calculation. It can also be 
seen in the Figure that there are considerably more gluons produced than 
quarks.

In order to get a better overall understanding, we present in 
the following two Figs. ((5) and (6)) the 
source term integrated over positive times, {\it i.e}:

\be
T=\frac{dW}{2d^3xd^3k}.
\ee

The time-integrated source terms can be regarded as a measure 
for the net-production of particles in an infinitesimal volume around any 
given point in the phase-space. It does not show the oscillatory behavior 
which gives a totally different picture for different times.
For our choice of the field (Eq. (\ref{as})) $T$ takes the form:

\bea
\frac{dW_{q\bar{q}}}{2d^3x d^3k}=
16\frac{\alpha_S}{(2\pi)^2}
(A_{in})^2
(\frac{t_0}{1+4(k_T)^2\cosh^2(y)t_0^2})^2
\frac{1}{\cosh^2(y)}
\eea

and

\bea
\frac{dW_{gg}}{2d^3xd^3k}
=
3\frac{\alpha_S}
      {(2\pi)^4}(A_{in})^2
 (-16(\frac{t_0}
          {1+4(k_T)^2\cosh^2(y)t_0^2})^2
   [3+\frac{1}
            {\cosh^2(y)}]
~\nonumber \\
  +3(A_{in})^2\frac{g^2}
                   {(k_T)^2\cosh^2(y)}
   (\frac{t_0}
         {1+(k_T)^2\cosh^2(y)t_0^2})^2)
\eea

respectively.

The decay behavior with the transverse momentum observed in Fig. 
(5) is mostly due to the choice of the field. Only in the 
second 
contribution to the gluon source term there is already a factor $1/(k^0)^2$ 
present in the general formula. Choosing e.g. a Gaussian form for the field, 
would also have resulted in a Gaussian decrease of the production rate with 
the transverse momentum of the particles.
As the momentum structure of the general equations is mostly based on the 
$k^0$-component, the origin for the typical rapidity behavior is mainly the 
same as for the behavior for changing transverse momentum, see Fig. 
(6).

In the last plot Fig. (7), we plot the ratio of the source terms 
for quark-antiquark pairs and gluon pairs:

\be
R=\frac{dW_{q\bar{q}}}{2d^3x d^3k}/\frac{dW_{gg}}{d^3xd^3k}.
\ee

It can now be seen directly from Fig. (7) that there are less 
quarks produced than gluons.
We observe for this model field that a stronger coupling, a stronger 
chromofield and/or a slowlier varying field emphasize the gluon-pair  
production even more in comparison to the production of 
$q\bar{q}$ pairs.  

%%%%%%%%%%%%%%%%%%%%%%%%%%%%%%%%%%%%%%%%%%%%%%%%%%%%%%%%%%%%%%%%%%%%%%%%%%%%%%%

\section{Conclusion}

We have derived the source term for quark-antiquark and gluon pair production
via vacuum polarization in the presence of a space-time dependent chromofield 
in the order $S^{(1)}$. We have used the background field method of QCD for
this purpose. The production of quark-antiquark pairs from a Yang-Mills 
field is very similar to the production of electron-positron pairs 
from a Maxwell field in QED. To the order $S^{(1)}$ the source 
term for gluon pairs in the presence of a general space-time dependent 
chromofield consists of three contributions, each of a different order 
in the coupling constant $g$.

To obtain an insight into the source terms, we have derived them for a purely 
time dependent chromofield and have presented the results for an exponentially
decaying field. In comparison to the arbitrary space-time 
dependent field, the equations obtained in this case are considerably simpler 
but still too complicated as to interpret them at first sight.
For this reason, not for giving a realistic physical prediction, we 
evaluated the source term in the purely time-dependent case for a given form 
of the chromofield. It was chosen, so that analytical solutions were 
possible, but still it is motivated by physical arguments. The results thus
obtained have been plotted and discussed.

We observe that the source term not only creates partons from the field but
also destroys them to produce field.
One can recognize periodic fluctuations of the production rate with time which
become more rapid for higher particle energies. This behavior is inherent to 
the general time-dependent formulas and is not connected to a special choice 
of the chromofield. 
This feature may play an important role in the evolution of the quark-gluon 
plasma. The mere existence of the field might complicate the transport 
studies. For example, one usually solves the hydrodynamic evolution equation 
\cite{bjorken}  when the plasma is in local thermal equilibrium. The presence 
of the chromofield results in the necessity to solve chromohydrodynamic 
evolution equations, thus one needs to know how fast the field decays. The 
description of all processes will become substantially easier if the field 
strength is already negligible before equilibration sets in.
In all plots it can be seen that there are considerably 
more gluon pairs than quark-antiquark pairs produced by this mechanism. 

Finally, we discuss the range of validity of the source terms derived in this 
paper. Being perturbative, our result is only applicable in situations, where 
certain conditions are satisfied.
Firstly, in the framework of our approach we can merely properly describe the 
production of partons with momentum $p$ larger than the field strength
$A$ multiplied 
by the coupling constant $g$. This means, if the field is very strong and $gA$ 
is truely large, our calculation is not valid, even if the coupling $g$ is 
weak. This is similarly true for the Schwinger mechanism for pair production
in QED in the presence of a weak field (see Eq. 6.33 in \cite{schwinger}).
Although QED is a weak coupling theory the perturbative calculation of 
Schwinger is applicable only in the limit of small $eA$. However, for a very 
small coupling, a truely large background field has got to be present before 
this condition is violated for reasonable momenta $p$.
Secondly, for the perturbative method to be applicable $gA$ has got to be 
larger than the non-perturbative scale $\L$. For the condition $gA<\L$ the
perturbative calculation is not valid and parton production
has to be computed via non-perturbative methods, which is unfortunately
non-trivial. So, the region in which our perturbative approach is applicable
is given by : $p>gA>\L$.

Let us consider a realistic situation in a heavy-ion collision at LHC (Pb-Pb 
at $\sqrt{s}$= 5.5 TeV). For the energy density formed in this
experiment one expects $\epsilon\sim$ 1000 GeV/fm$^3$ 
\cite{nayakk,keijo,raju} and for
the strong coupling constant $\alpha_s \sim$ 0.15 \cite{nayakk}. Assuming
that all the energy density $\epsilon$ is deposited in the field 
sector we can get a rough estimate for $gA$ ($E \sim \sqrt{2 \epsilon}$
and $gA^2 \sim E$) to be: $gA \sim$ 2 GeV. 
Hence, initially,  a perturbative treatment
might be applicable for the production of partons whose momentum is greater
than 2 GeV or so at LHC. At RHIC (Au-Au collisions at $\sqrt{s}$
= 200 GeV) the typical energy density and coupling contstant would
be $\epsilon\sim$ 50 GeV/fm$^3$ and $\alpha_S\sim$ 0.33 respectively. 
For this situation
one would expect roughly $gA \sim$ 1 GeV. So, for these input parameters 
at RHIC and LHC, the minimum
momentum above which one would apply perturbative calculations
is about 1 and 2 GeV respectively.
It is interesting to note that
these values are more or less equal to the minium momentum cut-offs used 
to compute minijets by using pQCD at RHIC and LHC \cite{muller,keijo}.
We have to mention that even if $gA$ is small the contributions of both 
diagrams have to be taken in order to maintain exact gauge invariance.

It is clear from the above arguments that during the very early stage of
heavy-ion collisions our perturbative calculation is applicable
for the production of partons having a momentum greater than $gA$.
Begining from such values of the field strength, it is now necessary 
to investigate how the field 
decays owing to all effects such as particle production from the field, 
acceleration of the partons by the field, precession of the color charge in 
the presence of the field, collision among the partons, and expansion of the 
system. As time progresses the field strength decreases due to the effects 
mentioned above. However, as long as $gA$ is greater than $\L$, our 
perturbative method is applicable. The decrease of 
the product $gA$ with time allows for the description of partons with less 
minimum momentum than right at the start of the URHIC. The picture breaks 
down however, when $gA$ reaches $\L$. In that case non-perturbative 
calculations for parton production are needed, which is beyond the
scope of this paper. Unfortunately, when trying to perform non-perturbative 
calculations one always encounters many more obstacles.

How rapidly the boundary of applicability of our picture is reached and how the
quark-gluon plasma evolves (up to a time where the condition $gA>\L$ is
violated) with our treatment of parton production
from the space-time dependent chromofield, must - preferrably - be determined 
from a self-consisitent transport calculation with all the above effects 
taken into account. This project will involve a considerable amount of 
numerical work and will be undertaken in future.

%%%%%%%%%%%%%%%%%%%%%%%%%%%%%%%%%%%%%%%%%%%%%%%%%%%%%%%%%%%%%%%%%%%%%%%%%%%%%%%

\section*{Acknowledgments}

The authors want to thank Larry McLerran for his critical comments and 
suggestions. 
Furthermore, the authors would like to thank
Stefan Hofman, Chung-Wen Kao, Alexander Krasnitz, Dirk Rischke,  
Raju Venugopalan, and Qun Wang for their helpful discussions.
D.D.D. would like to thank the Graduiertenf\"orderung des Landes 
Hessen for financial support.
G.C.N. would like to thank the Alexander von Humboldt Foundation for 
financial support.

%%%%%%%%%%%%%%%%%%%%%%%%%%%%%%%%%%%%%%%%%%%%%%%%%%%%%%%%%%%%%%%%%%%%%%%%%%%%%%%

\section*{Appendix}

\subsection*{Gluon Source Term}

We start from definition (\ref{D1A1A})
and introduce the expression (\ref{ampy}) and subsequently Eq.(\ref{V1A}) to 
obtain:

\bea
W_{1A,1A}
=\frac{g^2}{32(2\pi)^2}
\int\frac{d^3k_1}{k_1^0}\frac{d^3k_2}{k_2^0}d^4Kd^4K'
\delta^{(4)}(K-k_1-k_2)\delta^{(4)}(K'-K)
A^{a\mu}(K)A^{*a'\mu'}(K')
~\nonumber \\
f^{abd}f^{a'bd}
[-2g_{\mu\rho}K_{\nu}+g_{\nu\rho}(k_1-k_2)_{\mu}+2g_{\mu\nu}K_{\rho}]
[-2g_{\mu'\rho}K'_{\nu}+g_{\nu\rho}(k'_1-k'_2)_{\mu'}+2g_{\mu'\nu}K'_{\rho}]
\eea

Now we make use of the Fourier transform of Eq.(\ref{fourier}). In the next 
steps, we first carry out the integration over $d^4K'$ and then over $d^4K$:

\bea
W_{1A,1A}
=\frac{3g^2}{8(2\pi)^6}
\int\frac{d^3k_1}{k_1^0}\frac{d^3k_2}{k_2^0}d^4xd^4x'
e^{i(k_1+k_2)\cdot(x-x')}
A^{a\mu}(x)A^{a\mu'}(x')
~\nonumber \\
~[2g_{\mu\mu'}(k_1+k_2)^2
 -2(k_1+k_2)_{\mu}(k_1+k_2)_{\mu'}
 +(k_1-k_2)_{\mu}(k_1-k_2)_{\mu'}]
\eea

In the previous step, use has been made of Eq.(\ref{ff}). Now, we extract the 
expression (\ref{dW1A1A_}) for the source term.
From there, we proceed by replacing the occurrences of $k_2$ by derivatives 
with respect to $i(x-x')$:

\bea
\frac{dW_{1A,1A}}{d^4xd^3k}
=\frac{3g^2}{8(2\pi)^6k^0}
\int d^4x'\frac{d^3k_2}{k_2^0}
e^{ik\cdot(x-x')}
A^{a\mu}(x)A^{a\mu'}(x')
[2g_{\mu\mu'}(k+\frac{\partial}{i\partial(x-x')})^2
~\nonumber \\
 -2(k+\frac{\partial}{i\partial(x-x')})_{\mu}
   (k+\frac{\partial}{i\partial(x-x')})_{\mu'}
 +(k-\frac{\partial}{i\partial(x-x')})_{\mu}
  (k-\frac{\partial}{i\partial(x-x')})_{\mu'}]
e^{ik_2\cdot(x-x')}
\eea

Now the integration over $d^3k_2$ can be carried out with the help of 
Eq. (\ref{int1})
In the end, we explicitly calculate the derivatives and obtain 
Eq.(\ref{dW1A1A}).

Now we carry out the same procedure for the other contributions. First, we
calculate the probability of the mixed term. Starting from the definition 
Eq.(\ref{W1A2A})
we use the expressions (\ref{ampy}) and (\ref{ampx}) and get by subsequent 
introduction of Eqs. (\ref{V1A}) and (\ref{V2A}):

\bea
W_{1A,2A}=W_{2A,1A}=
\frac{-ig^3}{64(2\pi)^4}
\int\frac{d^3k_1}{k_1^0}\frac{d^3k_2}{k_2^0}d^4k_3 d^4k_4d^4K
\delta^{(4)}(K-k_3-k_4)\delta^{(4)}(K-k_1-k_2)
~\nonumber \\
A^{a\mu}(k_3)A^{c\lambda}(k_4)A^{*a'\mu'}(K)
[-2g_{\mu'\rho}K_{\nu}+
g_{\nu\rho}(k_1-k_2)_{\mu'}+
2g_{\mu'\nu}K_{\rho}]
~\nonumber \\
~[f^{a'bd}f^{abx}f^{xcd}
(g_{\mu\lambda}g_{\nu\rho}-g_{\mu\rho}g_{\nu\lambda}+g_{\mu\nu}g_{\lambda\rho})
~\nonumber \\
+f^{a'bd}f^{adx}f^{xbc}
(g_{\mu\nu}g_{\lambda\rho}-g_{\mu\lambda}g_{\nu\rho}-g_{\mu\rho}g_{\nu\lambda})
~\nonumber \\
+f^{a'bd}f^{acx}f^{xbd}
(g_{\mu\nu}g_{\lambda\rho}-g_{\mu\rho}g_{\nu\lambda})].
\eea

Now, we contract all possible color and Lorentz indices and make use of the 
Fourier transform (\ref{fourier}). From there the source term (\ref{dW2A1A_})
can be extracted.
In the next step, the $d^4K$ and the $d^4k_4$ integration are carried out, 
followed by the $d^4k_3$ integration:

\bea
\frac{dW_{2A,1A}}{d^4xd^3k}
=\frac{-3ig^3}{8(2\pi)^{10}}
\int\frac{d^3k_2}{k^0k_2^0}d^4k_3d^4x_3d^4x_4
e^{i[k_3\cdot (x_3-x_4)+(k+k_2)\cdot (x_4-x)]}
~\nonumber \\
A^{a\mu}(x_3)A^{c\lambda}(x_4)A^{a'\mu'}(x)
f^{a'ac}(k+k_2)_{\lambda}g_{\mu\mu'}
~\nonumber \\
=\frac{-3ig^3}{8(2\pi)^{10}}
\int\frac{d^3k_2}{k^0k_2^0}d^4x_3d^4x_4
(2\pi)^4\delta^{(4)}(x_3-x_4)
e^{i(k+k_2)\cdot (x_4-x)}
~\nonumber \\
A^{a\mu}(x_3)A^{c\lambda}(x_4)A^{a'\mu'}(x)
f^{a'ac}(k+k_2)_{\lambda}g_{\mu\mu'}.
\eea

The new Dirac $\delta$ distribution can be used to perform one of the 
integrations over $d^4x$:

\bea
\frac{dW_{2A,1A}}{d^4xd^3k}
=\frac{-3ig^3}{8(2\pi)^6}
\int\frac{d^3k_2}{k^0k_2^0}d^4x_3
e^{i(k+k_2)\cdot (x_3-x)}
A^{a\mu}(x_3)A^{c\lambda}(x_3)A^{a'\mu'}(x)
f^{a'ac}(k+k_2)_{\lambda}g_{\mu\mu'}.
\eea

Now, $k_2$ has got to be replaced by a derivative with respect to $i(x_3-x)$
in order to enable the use of Eq. (\ref{int1}) as a menas to obtain 
Eq.(\ref{dW2A1A}).

The source term of the last contribution remains to be calculated. 
We receive by introducing Eq. (\ref{V2A}) into Eq.(\ref{ampx}) and all that 
into Eq. (\ref{W2A2A}):

\bea
W_{2A,2A}
=\frac{g^4}{128(2\pi)^6}
\int\frac{d^3k_1}{k_1^0}\frac{d^3k_2}{k_2^0}d^4k_3 d^4k_4d^4k'_3 d^4k'_4
~\nonumber \\
\delta^{(4)}(k_1+k_2-k_3-k_4)\delta^{(4)}(k_3+k_4-k'_3-k'_4)
%~\nonumber \\
A^{a\mu}(k_3)A^{c\lambda}(k_4)A^{*a'\mu'}(k'_3)A^{*c'\lambda'}(k'_4)
~\nonumber \\
~[f^{abx}f^{xcd}
(g_{\mu\lambda}g_{\nu\rho}-g_{\mu\rho}g_{\nu\lambda}+g_{\mu\nu}g_{\lambda\rho})
%~\nonumber \\
+f^{adx}f^{xbc}
(g_{\mu\nu}g_{\lambda\rho}-g_{\mu\lambda}g_{\nu\rho}-g_{\mu\rho}g_{\nu\lambda})
~\nonumber \\
+f^{acx}f^{xbd}
(g_{\mu\nu}g_{\lambda\rho}-g_{\mu\rho}g_{\nu\lambda})]\times
%~\nonumber \\
~[f^{a'bx'}f^{x'c'd}
(g_{\mu'\lambda'}g_{\nu\rho}-g_{\mu'\rho}g_{\nu\lambda'}+g_{\mu'\nu}g_{\lambda'\rho})
~\nonumber \\
+f^{a'dx'}f^{x'bc'}
(g_{\mu'\nu}g_{\lambda'\rho}-g_{\mu'\lambda'}g_{\nu\rho}-g_{\mu'\rho}g_{\nu\lambda'})
%~\nonumber \\
+f^{a'c'x'}f^{x'bd}
(g_{\mu'\nu}g_{\lambda'\rho}-g_{\mu'\rho}g_{\nu\lambda'})]
\eea

Now, all possible indices have to be contracted:

\bea
W_{2A,2A}
=\frac{g^4}{64(2\pi)^{14}}
\int\frac{d^3k_1}{k_1^0}\frac{d^3k_2}{k_2^0}
d^4k_3 d^4k_4d^4k'_3 d^4k'_4d^4x_3d^4x_4d^4x'_3d^4x'_4
e^{i[k_3\cdot x_3+k_4\cdot x_4-k'_3\cdot x'_3-k'_4\cdot x'_4]}
~\nonumber \\
\delta^{(4)}(k_1+k_2-k_3-k_4)\delta^{(4)}(k_3+k_4-k'_3-k'_4)
A^{a\mu}(x_3)A^{c\lambda}(x_4)A^{a'\mu'}(x'_3)A^{c'\lambda'}(x'_4)
~\nonumber \\
~[2g_{\mu\lambda}g_{\mu'\lambda'}(f^{abx}f^{xcd}+f^{adx}f^{xcb})
                                 (f^{a'bx'}f^{x'c'd}+f^{a'dx'}f^{x'c'b})
 +24g_{\mu\mu'}g_{\lambda\lambda'}f^{acx}f^{a'c'x}]
\eea

From this expression we get the source term (\ref{dW2A2A_}).
In the following steps the integrations over all $d^4k$s and afterwards over 
$d^4x_4$ and $d^4x'_4$ are performed: 

\bea
\frac{dW_{2A,2A}}{d^4xd^3k}
=\frac{g^4}{32(2\pi)^{14}}
\int\frac{d^3k_2}{k^0k_2^0}
d^4k_3d^4k'_3d^4x_4d^4x'_3d^4x'_4
e^{i[k_3\cdot x+(k+k_2-k_3)\cdot x_4
    -k'_3\cdot x'_3-(k+k_2-k'_3)\cdot x'_4]}
~\nonumber \\
A^{a\mu}(x)A^{c\lambda}(x_4)A^{a'\mu'}(x'_3)A^{c'\lambda'}(x'_4)
~\nonumber \\
~[g_{\mu\lambda}g_{\mu'\lambda'}(f^{abx}f^{xcd}+f^{adx}f^{xcb})
                                 (f^{a'bx'}f^{x'c'd}+f^{a'dx'}f^{x'c'b})
 +12g_{\mu\mu'}g_{\lambda\lambda'}f^{acx}f^{a'c'x}]
~\nonumber \\
=\frac{g^4}{32(2\pi)^{14}}
\int\frac{d^3k_2}{k^0k_2^0}
d^4x_4d^4x'_3d^4x'_4
e^{i(k+k_2)\cdot(x_4-x'_4)}
(2\pi)^4\delta^{(4)}(x-x_4)(2\pi)^4\delta^{(4)}(x'_3-x'_4)
~\nonumber \\
A^{a\mu}(x)A^{c\lambda}(x_4)A^{a'\mu'}(x'_3)A^{c'\lambda'}(x'_4)
~\nonumber \\
~[g_{\mu\lambda}g_{\mu'\lambda'}(f^{abx}f^{xcd}+f^{adx}f^{xcb})
                                 (f^{a'bx'}f^{x'c'd}+f^{a'dx'}f^{x'c'b})
 +12g_{\mu\mu'}g_{\lambda\lambda'}f^{acx}f^{a'c'x}]
~\nonumber \\
=\frac{g^4}{32(2\pi)^6}
\int\frac{d^3k_2}{k^0k_2^0}
d^4x'_3
e^{i(k+k_2)\cdot(x-x'_3)}
A^{a\mu}(x)A^{c\lambda}(x)A^{a'\mu'}(x'_3)A^{c'\lambda'}(x'_3)
~\nonumber \\
~[g_{\mu\lambda}g_{\mu'\lambda'}(f^{abx}f^{xcd}+f^{adx}f^{xcb})
                                 (f^{a'bx'}f^{x'c'd}+f^{a'dx'}f^{x'c'b})
 +12g_{\mu\mu'}g_{\lambda\lambda'}f^{acx}f^{a'c'x}]
\eea

For the integration over $d^3k_2$ we again make use of Eq. (\ref{int1}) and 
obtain the final expression Eq.(\ref{dW2A2A}).

%  %  %  %  %  %  %  %  %  %  %  %  %  %  %  %  %  %  %  %  %  %  %  %  %  %  %

\subsection*{Gluon source term for the time-dependent case}

In the case were the $A$ fields are only time dependent, we receive from 
Eq. (\ref{dW1A1A_}) by integrating over the spatial coordinates:

\bea
\frac{dW_{1A,1A}}{d^4xd^3k}
=\frac{3g^2}{8(2\pi)^3}
\int\frac{d^3k_2}{k^0k_2^0}dt'
\delta^{(3)}(\vec k+\vec k_2)
e^{i[(k+k_2)\cdot x-(k^0+k_2^0)t']}
A^{a\mu}(t)A^{a\mu'}(t')
~\nonumber \\
~[2g_{\mu\mu'}(k+k_2)^2
 -2(k+k_2)_{\mu}(k+k_2)_{\mu'}
 +(k-k_2)_{\mu}(k-k_2)_{\mu'}]
\eea

By eliminating all Dirac $\delta$ distributions one obtains:

\bea
\frac{dW_{1A,1A}}{d^4xd^3k}
=\frac{3g^2}{8(2\pi)^3 4(k^0)^2}
\int dt'
e^{2ik^0(t-t')}
~\nonumber \\
~4[-2(\vec A^{a}(t)\cdot\vec A^{a}(t'))(k^0)^2
 +(\vec A^a(t)\cdot\vec k)(\vec A^a(t')\cdot\vec k)],
\eea

where the result (\ref{dW1A1At}) can be obtained incorporating Fourier 
transforms.

As before, we integrate in Eq. (\ref{dW2A1A_}) over all spatial coordinates 
and subsequently integrate over all $d^4k$s followed by $dt_4$:

\bea
\frac{dW_{2A,1A}}{d^4xd^3k}
=\frac{-3ig^3}{8(2\pi)^4}
\int\frac{d^3k_2}{k^0k_2^0}d^4k_3 d^4k_4d^4Kd^4t_3d^4t_4
\delta^{(4)}(K-k_3-k_4)\delta^{(4)}(K-k-k_2)
~\nonumber \\
\delta^{(3)}(\vec k_3)\delta^{(3)}(\vec k_4)
e^{i[k_3^0t_3+k_4^0t_4-K\cdot x]}
A^{a\mu}(t_3)A^{c\lambda}(t_4)A^{a'\mu'}(t)
f^{a'ac}K_{\lambda}g_{\mu\mu'}
~\nonumber \\
=\frac{-3ig^3}{8(2\pi)^4}
\int\frac{d^3k_2}{k^0k_2^0}d^4k_3d^4t_3d^4t_4
\delta^{(3)}(\vec k_3)\delta^{(3)}(\vec k+\vec k_2)
e^{i[k_3^0t_3+(k^0+k_2^0-k_3^0)t_4-(k+k_2)\cdot x]}
~\nonumber \\
A^{a\mu}(t_3)A^{c\lambda}(t_4)A^{a'\mu'}(t)
f^{a'ac}(k+k_2)_{\lambda}g_{\mu\mu'}
~\nonumber \\
=\frac{-3ig^3}{8(2\pi)^4}
\int\frac{d^3k_2}{k^0k_2^0}d^4t_3d^4t_4
\delta^{(3)}(\vec k+\vec k_2)(2\pi)\delta(t_3-t_4)
e^{i[(k^0+k_2^0)t_4-(k+k_2)\cdot x]}
~\nonumber \\
A^{a\mu}(t_3)A^{c\lambda}(t_4)A^{a'\mu'}(t)
f^{a'ac}(k+k_2)_{\lambda}g_{\mu\mu'}
~\nonumber \\
=\frac{-3ig^3}{8(2\pi)^3(k^0)^2}
\int dt_3
e^{2ik^0(t_3-t)}
A^{a\mu}(t_3)A^{c0}(t_3)A^{a'\mu'}(t)
f^{a'ac}2k^0g_{\mu\mu'}
~\nonumber \\
=\frac{-3ig^3}{4(2\pi)^3k^0}
\int dt_3
e^{2ik^0(t_3-t)}
A^{a\mu}(t_3)A^{c0}(t_3)A^{a'\mu'}(t)
f^{a'ac}g_{\mu\mu'}.
\eea

Starting from Eq. (\ref{dW2A2A_}), we carry out similar steps for the last 
gluon term:

\bea
\frac{dW_{2A,2A}}{d^4xd^3k}
=\frac{g^4}{32(2\pi)^5}
\int\frac{d^3k_2}{k^0k_2^0}
d^4k_3d^4k_4d^4k'_3 d^4k'_4dt_4dt'_3dt'_4
e^{i[k_3\cdot x+k^0_4t_4-k^{'0}_3t'_3-k^{'0}_4t'_4]}
~\nonumber \\
\delta^{(3)}(\vec k_4)\delta^{(3)}(\vec k'_3)\delta^{(3)}(\vec k'_4)
~\nonumber \\
\delta^{(4)}(k+k_2-k_3-k_4)\delta^{(4)}(k_3+k_4-k'_3-k'_4)
A^{a\mu}(t)A^{c\lambda}(t_4)A^{a'\mu'}(t'_3)A^{c'\lambda'}(t'_4)
~\nonumber \\
~[g_{\mu\lambda}g_{\mu'\lambda'}(f^{abx}f^{xcd}+f^{adx}f^{xcb})
                                 (f^{a'bx'}f^{x'c'd}+f^{a'dx'}f^{x'c'b})
 +12g_{\mu\mu'}g_{\lambda\lambda'}f^{acx}f^{a'c'x}]
~\nonumber \\
=\frac{g^4}{32(2\pi)^5}
\int\frac{d^3k_2}{k^0k_2^0}
d^4k_3d^4k'_3dt_4dt'_3dt'_4
e^{i[k_3\cdot x+(k^0+k^0_2-k^0_3)t_4-k'_3t'_3-(k^0+k^0_2-k^{'0}_3)t'_4]}
~\nonumber \\
\delta^{(3)}(\vec k+\vec k_2-\vec k_3)\delta^{(3)}(\vec k'_3)
\delta^{(3)}(\vec k+\vec k_2-\vec k'_3)
A^{a\mu}(t)A^{c\lambda}(t_4)A^{a'\mu'}(t'_3)A^{c'\lambda'}(t'_4)
~\nonumber \\
~[g_{\mu\lambda}g_{\mu'\lambda'}(f^{abx}f^{xcd}+f^{adx}f^{xcb})
                                 (f^{a'bx'}f^{x'c'd}+f^{a'dx'}f^{x'c'b})
 +12g_{\mu\mu'}g_{\lambda\lambda'}f^{acx}f^{a'c'x}]
~\nonumber \\
=\frac{g^4}{32(2\pi)^5}
\int\frac{d^3k_2}{k^0k_2^0}
dt_4dt'_3dt'_4
e^{i(k^0+k^0_2)(t_4-t'_4)}
~\nonumber \\
\delta^{(3)}(\vec k+\vec k_2)
(2\pi)\delta(t-t_4)(2\pi)\delta(t'_3-t'_4)
A^{a\mu}(t)A^{c\lambda}(t_4)A^{a'\mu'}(t'_3)A^{c'\lambda'}(t'_4)
~\nonumber \\
~[g_{\mu\lambda}g_{\mu'\lambda'}(f^{abx}f^{xcd}+f^{adx}f^{xcb})
                                 (f^{a'bx'}f^{x'c'd}+f^{a'dx'}f^{x'c'b})
 +12g_{\mu\mu'}g_{\lambda\lambda'}f^{acx}f^{a'c'x}]
~\nonumber \\
=\frac{g^4}{32(2\pi)^3(k^0)^2}
\int dt'_3
e^{2ik^0(t-t'_3)}
A^{a\mu}(t)A^{c\lambda}(t_3)A^{a'\mu'}(t'_3)A^{c'\lambda'}(t'_3)
~\nonumber \\
~[g_{\mu\lambda}g_{\mu'\lambda'}(f^{abx}f^{xcd}+f^{adx}f^{xcb})
                                 (f^{a'bx'}f^{x'c'd}+f^{a'dx'}f^{x'c'b})
 +12g_{\mu\mu'}g_{\lambda\lambda'}f^{acx}f^{a'c'x}].
\eea

%%%%%%%%%%%%%%%%%%%%%%%%%%%%%%%%%%%%%%%%%%%%%%%%%%%%%%%%%%%%%%%%%%%%%%%%%%%%%%%

\newpage

%%%%%%%%%%%%%%%%%%%%%%%%%%%%%%%%%%%%%%%%%%%%%%%%%%%%%%%%%%%%%%%%%%%%%%%%%%%%%%%
\begin{figure}[thb]
\begin{center}
\pfig{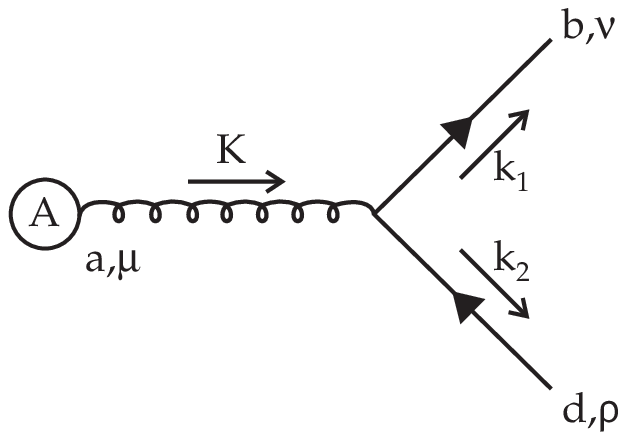}{12cm}
{Feynman diagram for the production of an quark-antiquark pair by 
coupling to the field $A$ once.}
\end{center}
\end{figure}
%%%%%%%%%%%%%%%%%%%%%%%%%%%%%%%%%%%%%%%%%%%%%%%%%%%%%%%%%%%%%%%%%%%%%%%%%%%%%%%
\begin{figure}[thb]
\begin{center}
\pfig{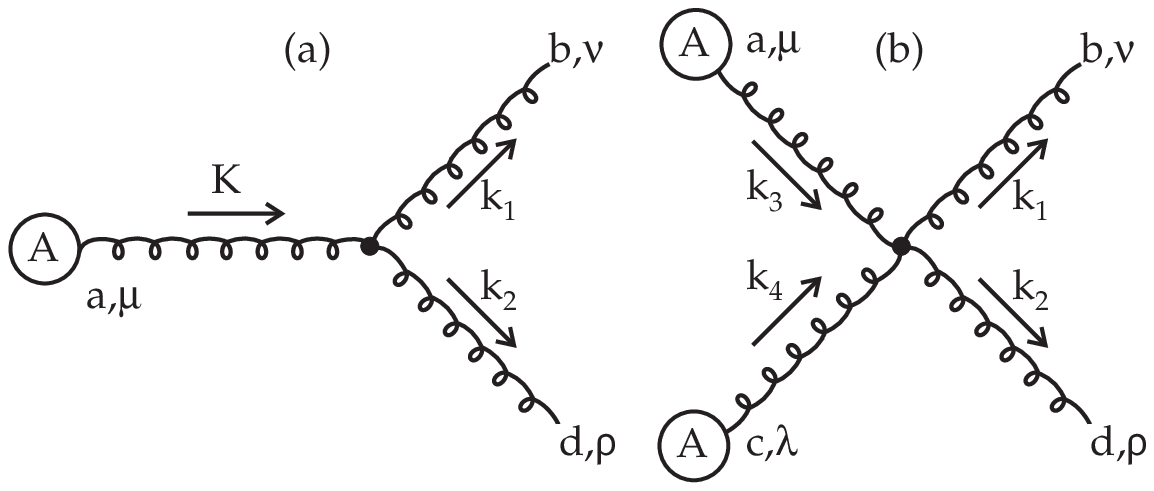}{12cm}
{Feynman diagrams for the production of two gluons by coupling to the 
field $A$ once (a) or twice (b).}
\end{center}
\end{figure}
%%%%%%%%%%%%%%%%%%%%%%%%%%%%%%%%%%%%%%%%%%%%%%%%%%%%%%%%%%%%%%%%%%%%%%%%%%%%%%%
\begin{figure}[thb]
\begin{center}
\pfig{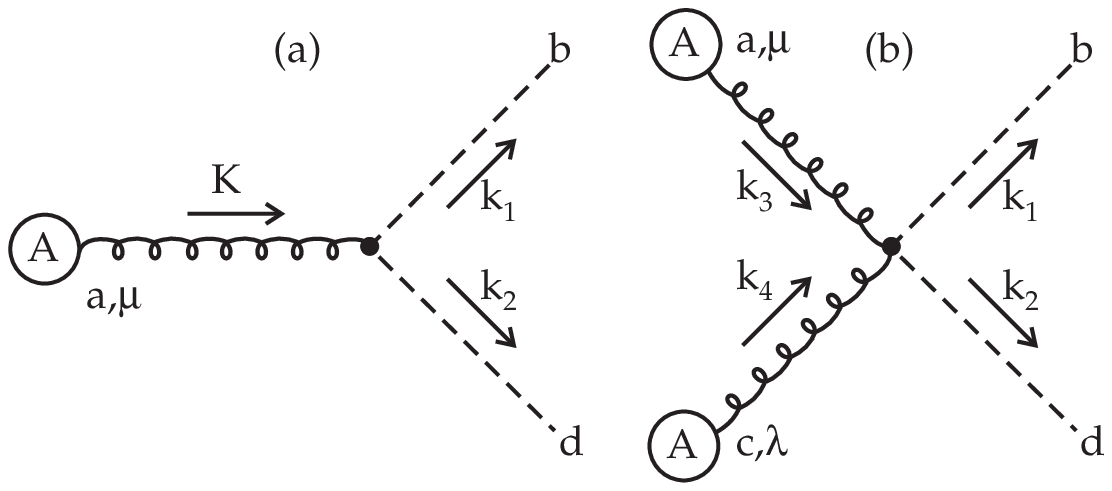}{12cm}
{Feynman diagrams for the ghosts, corresponding to the gluon vertices in 
Fig.(2).}
\end{center}
\end{figure}
%%%%%%%%%%%%%%%%%%%%%%%%%%%%%%%%%%%%%%%%%%%%%%%%%%%%%%%%%%%%%%%%%%%%%%%%%%%%%%%
\begin{figure}[thb]
\begin{center}
\pfig{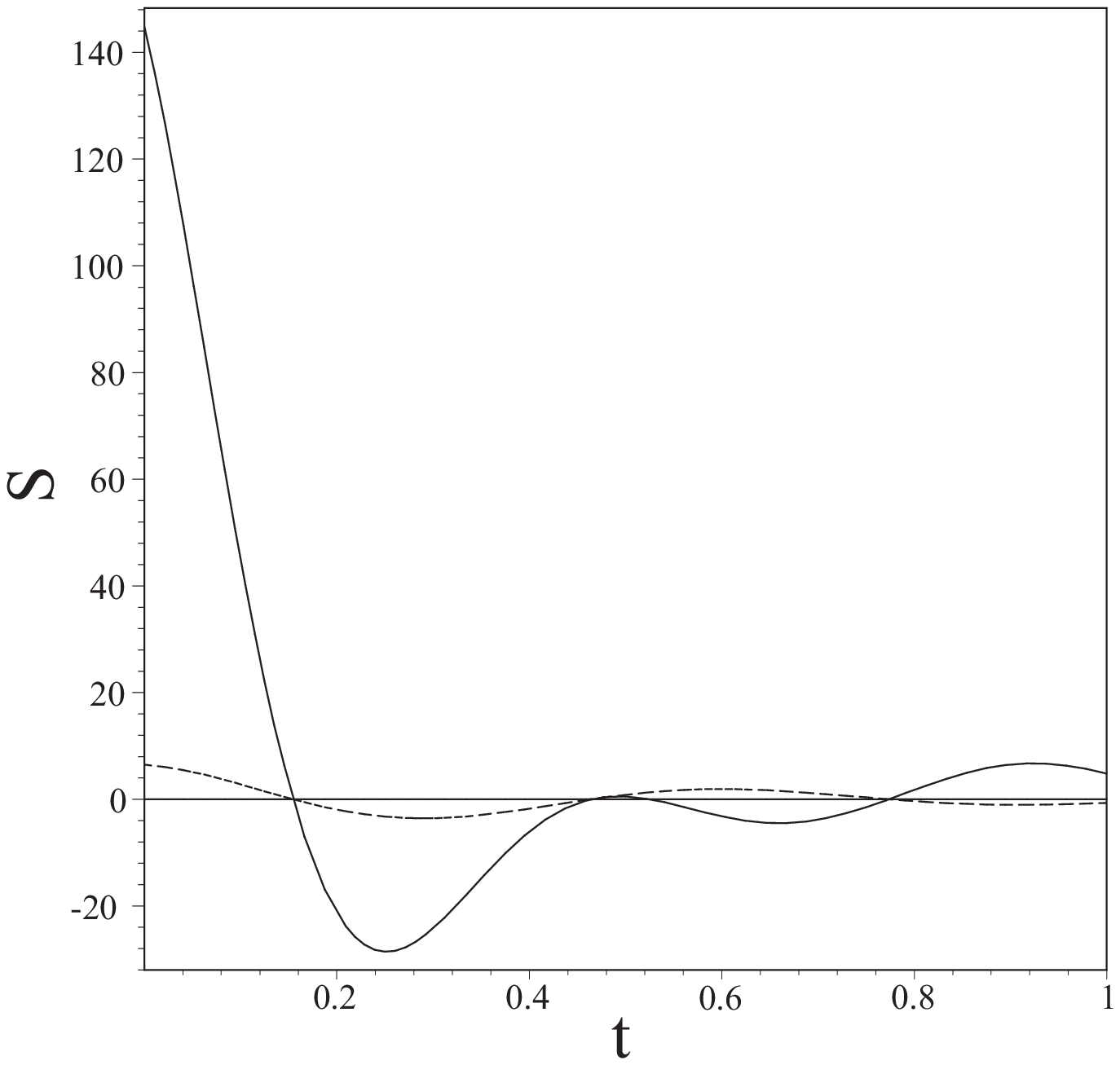}{12cm}
{Source terms $S$ in $MeV$ for the production of gluon pairs 
(solid line) and quark-antiquark pairs (dashed line) versus time $t$ in $fm$.}
\end{center}
\end{figure}
%%%%%%%%%%%%%%%%%%%%%%%%%%%%%%%%%%%%%%%%%%%%%%%%%%%%%%%%%%%%%%%%%%%%%%%%%%%%%%%
\begin{figure}[thb]
\begin{center}
\pfig{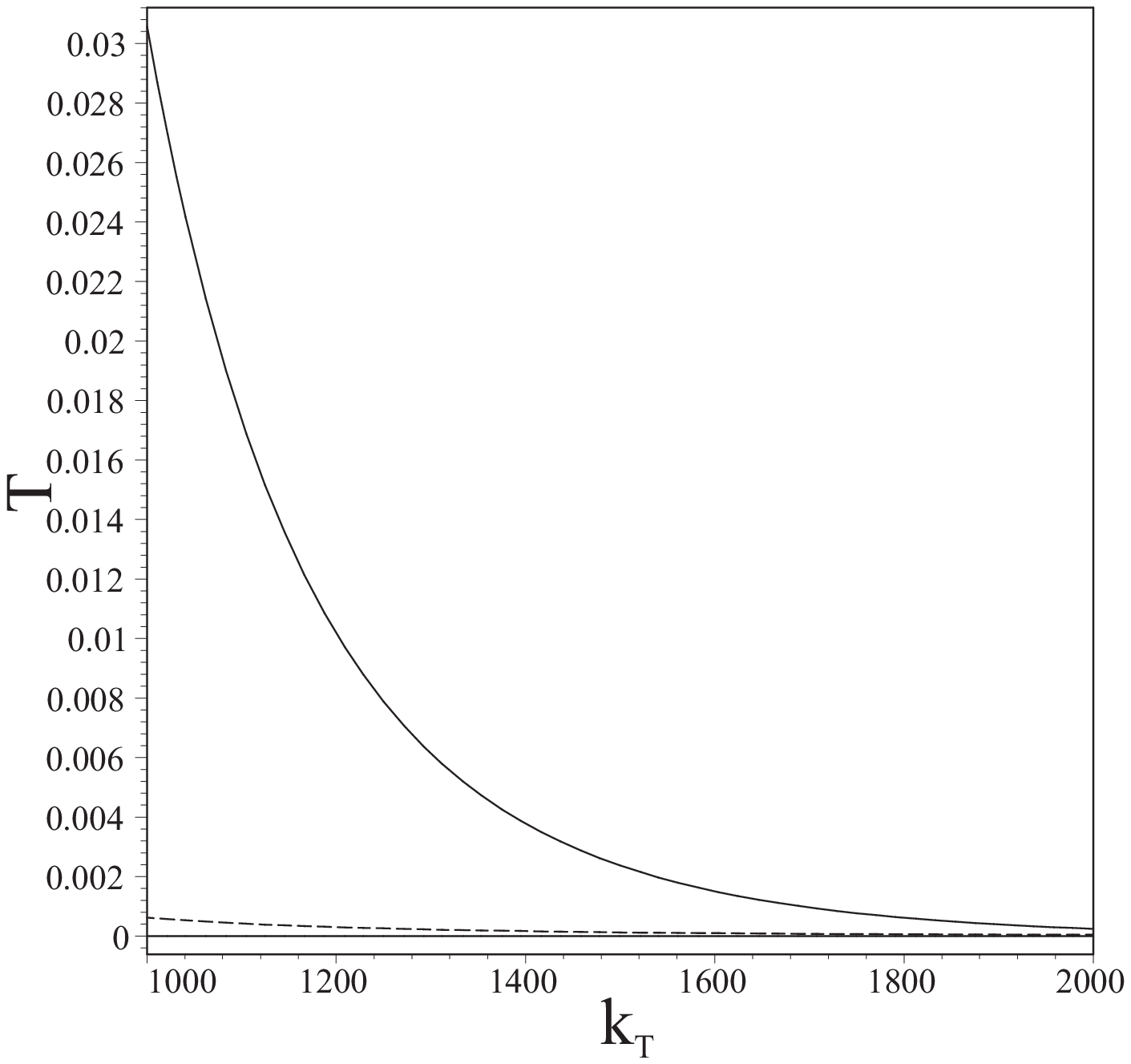}{12cm}
{The dimensionless time-integrated source terms $T$ for the production 
of gluon pairs (solid line) and quark-antiquark pairs (dashed line) versus 
transverse momentum $k_T$ in $MeV$.}
\end{center}
\end{figure}
%%%%%%%%%%%%%%%%%%%%%%%%%%%%%%%%%%%%%%%%%%%%%%%%%%%%%%%%%%%%%%%%%%%%%%%%%%%%%%%
\begin{figure}[thb]
\begin{center}
\pfig{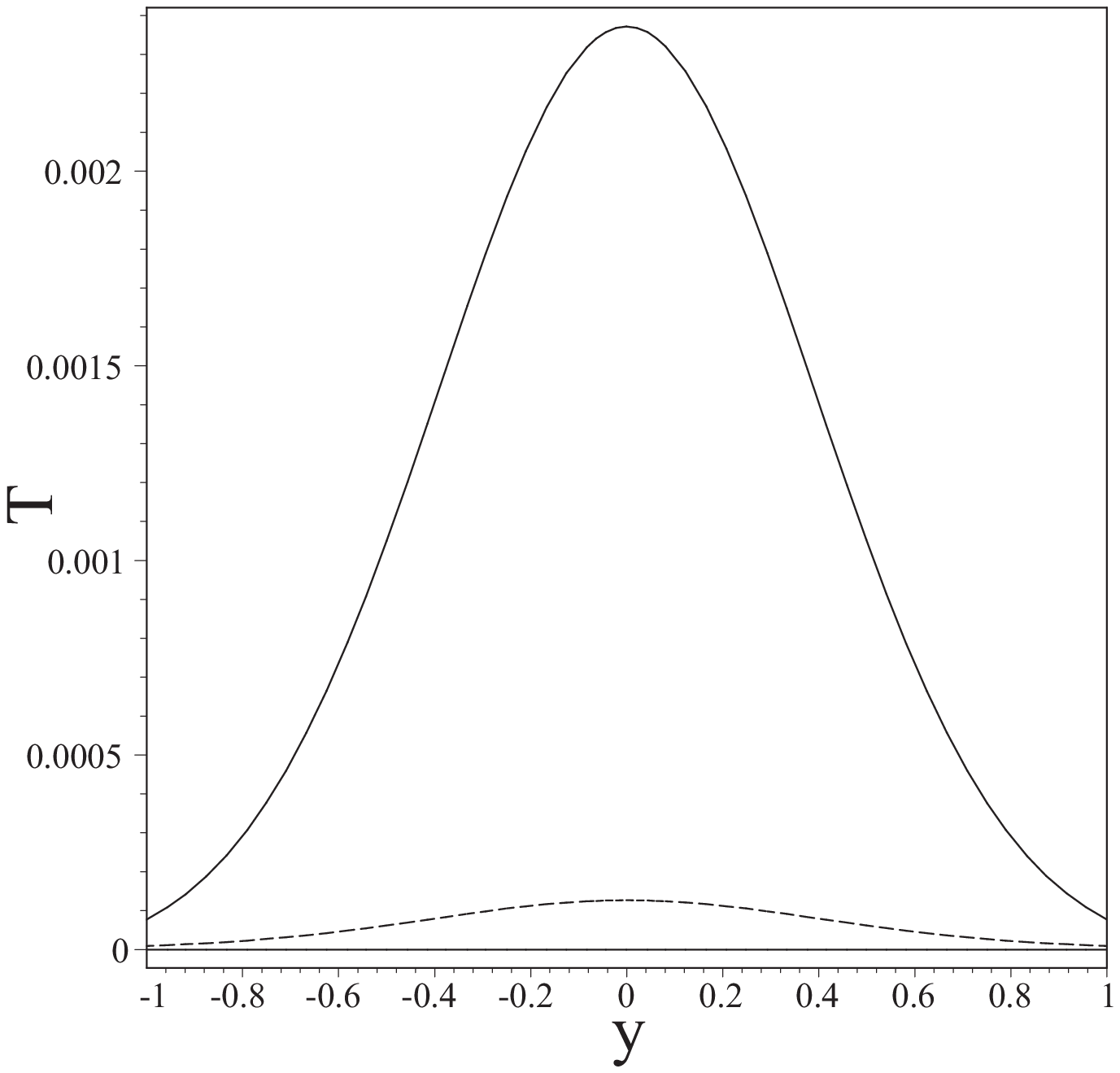}{12cm}
{The dimensionless time-integrated source terms $T$ for the production 
of gluon pairs (solid line) and quark-antiquark pairs (dashed line) versus 
rapidity $y$.}
\end{center}
\end{figure}
%%%%%%%%%%%%%%%%%%%%%%%%%%%%%%%%%%%%%%%%%%%%%%%%%%%%%%%%%%%%%%%%%%%%%%%%%%%%%%%
\begin{figure}[thb]
\begin{center}
\pfig{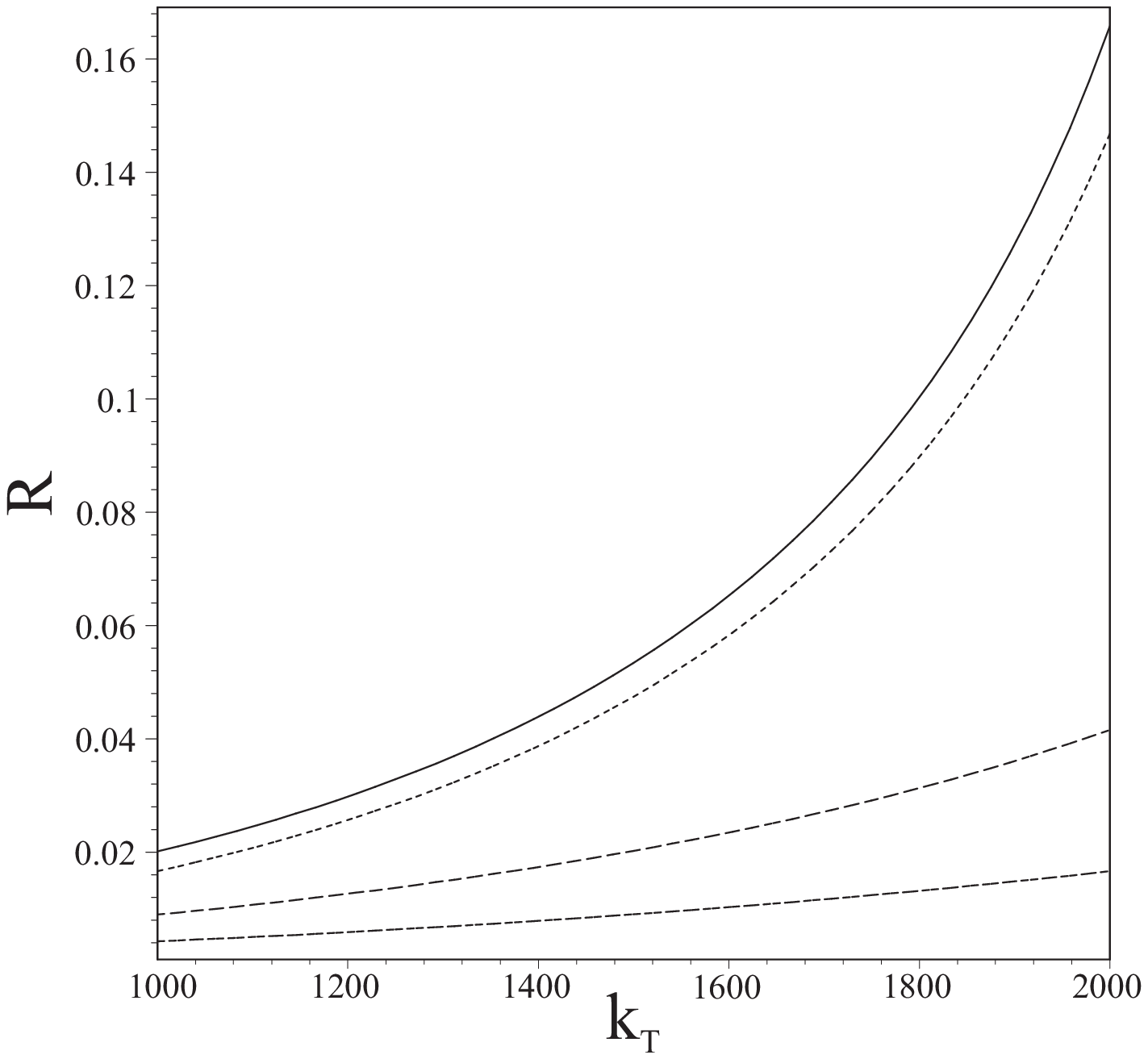}{12cm}
{Ratio $R$ of the time-integrated source terms for the production of 
gluon pairs and quark-antiquark pairs versus transverse momentum $k_T$ in 
$MeV$. From top to bottom we have first the graph for the standard parameters: 
coupling constant $\alpha_S=0.15$, initial field strength $A_{in}=1.5GeV$, and 
decay time $t_0=0.5fm$. In the subsequent graphs we increased the coupling 
constant to $\alpha_S=0.3$, the initial field strength to $A_{in}=3GeV$, and 
the decay time to $t_0=1fm$, respectively.}
\end{center}
\end{figure}
%%%%%%%%%%%%%%%%%%%%%%%%%%%%%%%%%%%%%%%%%%%%%%%%%%%%%%%%%%%%%%%%%%%%%%%%%%%%%%%


\begin{references}

\bibitem{qgp}
L.~.~Riccati, M.~.~Masera and E.~.~Vercellin,
Quark matter '99.  Proceedings, 14th International Conference on ultra-relativistic  nucleus nucleus collisions, QM'99, Torino, Italy, May
10-15, 1999,
in {\it NONE}
Nucl.\ Phys.\  {\bf A661} (1999) 1.

\bibitem{lattice} See, e.g., 
L. McLerran and B. Svetitsky, Phys. Rev. D24 (1981) 450; 
L. McLerran, Phys. Rev. D36 (1987) 3291;
R.V. Gavai, in {\it Quantum Fields on the Computer},
   ed. M. Creutz, (World Scientific, 1992), p. 51; 
F. Karsch and E. Laermann, Rep. Prog. Phys. {56} (1993) 1347; 
M. Oevers, F. Karsch, E. Laermann and P. Schmidt, in Proc. of Lattice '97:
   Nucl. Phys. Proc. Suppl. {63} (1998) 394.

\bibitem{soft}
F.~E.~Low, Phys.\ Rev.\  {\bf D12} (1975) 163;
S.~Nussinov, Phys.\ Rev.\ Lett.\  {\bf 34} (1975) 1286;
A. Karman, T. Matsui and B. Svetitsky, Phys. Rev. Lett.
{\bf 56}, 219 (1986) and references therein.

\bibitem{alll}
K.~Kajantie and T.~Matsui, Phys.\ Lett.\  {\bf B164} (1985) 373;
G.~Gatoff, A.~K.~Kerman and T.~Matsui, Phys.\ Rev.\  {\bf D36} (1987) 114;
A.~Bialas, W.~Czyz, A.~Dyrek and W.~Florkowski, Nucl.\ Phys.\  {\bf B296} 
(1988) 611;
B.~Banerjee, R.~S.~Bhalerao and V.~Ravishankar, Phys.\ Lett.\  {\bf B224} 
(1989) 16;
M.~Asakawa and T.~Matsui, Phys.\ Rev.\  {\bf D43} (1991) 2871;
K.~J.~Eskola and M.~Gyulassy, Phys.\ Rev.\  {\bf C47} (1993) 2329;
J.~M.~Eisenberg, Found.\ Phys.\  {\bf 27} (1997) 1213.

\bibitem{nayak} 
G. C. Nayak and V. Ravishankar, Phys. Rev.
D {\bf 55}, 6877 (1997); Phys. Rev. C {\bf 58}, 356 (1998).

\bibitem{sr}
C.~D.~Roberts and S.~M.~Schmidt, Prog. Part. Nucl. Phys.{\bf 45} Suppl.1:1-103,
2000, nucl-th/0005064, and references therein;
D.F. Litim and C. Manuel, Phys. Rev. Lett. {\bf 82}, 4981 (1999);
Nucl. Phys. {\bf B562} (1999) 237; Phys. Rev. {\bf D61}, 125004, (2000).

\bibitem{more}
Y.~Kluger, J.~M.~Eisenberg, B.~Svetitsky, F.~Cooper and E.~Mottola, 
Phys. Rev. Lett. {\bf 67} (1991) 2427;
F.~Cooper, J.~M.~Eisenberg, Y.~Kluger, E.~Mottola and B.~Svetitsky, 
Phys. Rev. {\bf D 48} (1993) 190;
J.~M.~Eisenberg, Phys. Rev. {\bf D 51} (1995) 1938;
F.~Cooper and E.~Mottola, Phys. Rev. {\bf D 36} (1987) 3114;
{\it ibid} {\bf D 40} (1989) 456;
T.~S.~Biro, H.~B.~Nielsen and J.~Knoll, Nucl. Phys. {\bf B 245} (1984) 449;
M.~Herrmann and J.~Knoll, Phys. Lett. {\bf B 234} (1990) 437;
D.~Boyanovsky, H.~J.~de Vega, R.~Holman, D.~S.~Lee and A.~Singh, 
Phys. Rev. {\bf D 51} (1995) 4419;
H.~Gies, Phys. Rev. {\bf D 61} (2000) 085021.

\bibitem{coll}
V.~D.~Barger and R.~J.~N.~Phillips, 
{\it Collider Physics} (Addison-Wesley Publishing Company, 1987).

\bibitem{lund}
B.~Andersson, G.~Gustafson, G.~Ingelman and T.~Sjostrand,
Phys.\ Rept.\  {\bf 97} (1983) 31;
B.~Andersson, G.~Gustafson and B.~Nilsson-Almqvist,
Nucl.\ Phys.\  {\bf B281} (1987) 289.

\bibitem{heinz} 
H-T Elze and U. Heinz, Phys. Rep. {\bf 183}, 81 (1989).

\bibitem{wong} 
S.K. Wong, Nuovo Cimento A {\bf 65}, 689 (1970).

\bibitem{schwinger} 
J. Schwinger, Phys. Rev. {\bf 82}, 664 (1951).

\bibitem{casher} 
A. Casher, H. Neuberger and S. Nussinov, Phys. Rev. 
D {\bf 20}, 179 (1979).

\bibitem{izju} 
C. Itzykson and J. Zuber, {\it Quantum Field Theory} (McGraw-Hill Inc., 1980); 
R. S. Bhalerao and V. Ravishankar, Phys. Lett. {\bf B409}, 38 (1997);
G. C. Nayak and W. Greiner, {\it hep-th/}0001009.

\bibitem{dewitt} 
B. S. DeWitt, Phys. Rev. {\bf 162}, 1195 and
1239 (1967); {\it in} Dynamic theory of groups and fields 
(Gordon and Breach, 1965).

\bibitem{thooft} 
G. 't Hooft, Nucl. Phys. {\bf B62}, 444 (1973).

\bibitem{mclerran}
L.~McLerran and R.~Venugopalan,Phys.\ Rev.\  {\bf D49}, 2233 (1994); 
{\it ibid} {\bf 49}, 3352 (1994);
A.~Kovner, L.~McLerran and H.~Weigert, Phys.\ Rev.\  {\bf D52} (1995) 3809;
{\it ibid}, {\bf 52} (1995) 6231.

\bibitem{raju}
A.~Krasnitz and R.~Venugopalan, Nucl.\ Phys.\  {\bf B557}, 237 (1999);
Phys.\ Rev.\ Lett.\  {\bf 84}, 4309 (2000).

\bibitem{maksur}
A.~Makhlin, Phys.\ Rev.\ C {\bf 63}, 044902 (2001) [hep-ph/0007300];
%%CITATION = HEP-PH 0007300;%%
A.~Makhlin and E.~Surdutovich, Phys.\ Rev.\  {\bf C58}, 389 (1998).

\bibitem{jana}
R.~Jackiw and V.~P.~Nair, Phys.\ Rev.\  {\bf D48}, 4991 (1993).

\bibitem{bodeker}
D.~Bodeker,
Phys.\ Lett.\  {\bf B426}, 351 (1998).

\bibitem{abst} 
M. Abramowitz and I. Stegun, {\it Handbook of Mathematical 
Functions} (Dover Publications, Inc, New York, 1972).

\bibitem{abbott} 
L. F. Abbott, Nucl. Phys. {\bf B185}, 189 (1981).

\bibitem{num} 
R. S. Bhalerao and G. C. Nayak, 
Phys. Rev.  {\bf C61}, 054907 (2000).

\bibitem{dng}
D.~D.~Dietrich, G.~C.~Nayak and W.~Greiner, hep-ph/0009178.

\bibitem{bjorken} 
J. D. Bjorken, Phys. Rev. D27 (1983) 140.

\bibitem{nayakk}
G. C. Nayak, A. Dumitru, L. McLerran, and W. Greiner, hep-ph/0001202, 
Nucl. Phys. A, 687 (2001) 457.

\bibitem{keijo}
K. J. Eskola, K. Kajantie, P. V. Ruuskanen, and K. Tuorninen, 
Nucl. Phys. B570 (2000) 379.

\bibitem{muller}
Yu. Kovchegov and A. H. Mueller, Nucl. Phys. B529 (1998) 451; A. H. Mueller,
Nucl. Phys. B572 (2000) 227.

\end{references}
\end{document}